\documentclass[12pt,preprint]{aastex}

\newbox\grsign \setbox\grsign=\hbox{$>$} \newdimen\grdimen \grdimen=\ht\grsign
\newbox\simlessbox \newbox\simgreatbox
\setbox\simgreatbox=\hbox{\raise.5ex\hbox{$>$}\llap
     {\lower.5ex\hbox{$\sim$}}}\ht1=\grdimen\dp1=0pt
\setbox\simlessbox=\hbox{\raise.5ex\hbox{$<$}\llap
     {\lower.5ex\hbox{$\sim$}}}\ht2=\grdimen\dp2=0pt
\def\simgreat{\mathrel{\copy\simgreatbox}}
\def\simless{\mathrel{\copy\simlessbox}}
\newbox\simppropto
\setbox\simppropto=\hbox{\raise.5ex\hbox{$\sim$}\llap
     {\lower.5ex\hbox{$\propto$}}}\ht2=\grdimen\dp2=0pt

\slugcomment{To Appear in The Astrophysical Journal}

\shorttitle{Population Synthesis in the Blue I}
\shortauthors{Schiavon et al.}

\begin{document}


\title{Population Synthesis in the Blue I. Synthesis of the Integrated
Spectrum of 47 Tucanae from its Color-Magnitude Diagram}


\author{Ricardo P. Schiavon,  S. M. Faber}
\affil{UCO/Lick Observatory, University of California, Santa Cruz, CA 95064.}
\email{ripisc,faber@ucolick.org}

\author{Bruno V. Castilho}
\affil{Laborat\'orio Nacional de Astrof\'\i sica, MCT,
CP 21, 37500-000 Itajub\'a, Brazil}
\email{bruno@lna.br}

\and
\author{James A. Rose}
\affil{Department of Physics and Astronomy, CB 3255, University of North
Carolina, Chapel Hill, NC 27599}
\email{jim@physics.unc.edu}




\begin{abstract}

We perform an empirical synthesis of the blue integrated spectrum
of the metal-rich globular cluster 47 Tucanae, based directly on
the color-magnitude diagram of the cluster coupled to a moderately
high-resolution spectral library.  Freed from any significant dependence
on theory, we are able to perform a fundamental test of the adequacy
of the spectral library and its associated stellar parameters.
Excellent fits are achieved for almost all absorption-line indices
studied, provided the computations are corrected for two limitations
of the spectral library, namely, the lack of a representative set of
metal-poor giants and the absence of CN-strong stars. The latter effect
is corrected by means of spectrum synthesis from model photospheres,
considering the abundance pattern of CN-strong and CN-normal stars.
We also need to perform a slight correction of the metallicity of
the cluster (by --0.05 dex in relation to the standard value {\it
[Fe/H]}=--0.7) in order to match the metal-line index measurements in
the cluster spectrum.  After these relatively small adjustments, the
overall spectral agreement is good. Good fits are achieved for $H\beta$,
$H\gamma$, $Mg\,b$, $<Fe>$, Ca4227 and Fe4383, and only $H\delta_F$ is
overpredicted. Thus, ages inferred from $H\delta_F$ are slightly older
than the ages based on the other Balmer lines, by $\sim$ 3 Gyrs. The
success of this exercise suggests that previous failures to synthesize the
spectrum of 47 Tuc must have arisen from inadequacies in the theoretical
evolutionary isochrones and/or luminosity functions. Such a possibility
is considered in a companion paper.

\end{abstract}




\section{Introduction}

The advent of 8-10m class telescopes and high-throughput spectrographs
is bringing about a boost of high quality spectroscopic data on distant
galaxies. In particular, blue absorption-line spectra of galaxies with
redshifts as high as $\sim$ 1.0 can now be obtained with moderately
high resolution and S/N, with relatively low integration times. From
such data it is possible to infer the luminosity-weighted ages and
metallicities of the stellar populations of distant galaxies, in order
to establish constraints on cosmological scenarios (see, for instance,
Kelson et al. 2001). However, such analyses are crucially dependent on
the availability of reliable high-resolution model integrated spectra
of stellar populations in the blue and UV spectral regions.

This is the first of a series of papers in which we will present a new
set of models for population synthesis in the blue/optical region. The
ultimate goal of this work is to estimate luminosity-weighted ages and
metal abundances of galaxies at intermediate-to-high redshifts, as part
of the DEEP Survey (Koo 1998). The useful spectral region for the study
of stellar populations of distant galaxies in the DEEP project is the
restframe interval $\lambda\lambda$3800--4500 {\AA}. Consequently, the
emphasis of this paper is on understanding the behavior of absorption-line
indices in the blue. This interval represents a compromise between the
green-red, which has well understood and calibrated spectral indices but
is too far to the red for high-$z$ work, and the UV, which is optimal
for high $z$ but still lacks comprehensive empirical spectral libraries
and reliable calibrations between spectral indices and stellar parameters.

In part because of the difficulties associated with the crowding of
atomic and molecular lines, the use of absorption lines in blue population
synthesis work is less highly developed than the use of lines at redder
wavelengths. Early work was pioneered by Rose (1985, 1994), who defined
a number of line and pseudocontinuum ratios that are sensitive to age
and metallicity. Only recently has the behavior of those indices as
a function of age and metallicity on the basis of stellar population
models been presented for the first time (Vazdekis 1999). The Lick index
system (Worthey et al. 1994) technically extends to wavelengths as blue
as $\sim$ 4100 {\AA}. However, the indices bluer than $\sim$ 4500 {\AA}
have been far less explored in galaxy work than indices such as $Mg\,b$,
$H\beta$ and $<Fe>$. The same can be said about recent extensions of
the Lick system to include higher-order Balmer lines as age indicators
(Jones \& Worthey 1995, Worthey \& Ottaviani 1997). More recently,
Vazdekis \& Arimoto (1999) proposed a new definition of the $H\gamma$
index which is very insensitive to metallicity, and thus considerably
improves the separation between age and metallicity effects. However,
it too has so far been studied in very few objects. Another important
issue insufficiently addressed in the literature is the influence of CN
and CH lines on the measurement of line indices in the blue, which may
be very important in view of the fact that giant ellipticals may have
nonsolar [C/Fe] (Trager et al. 1998).  Therefore it is fair to say that
our understanding of blue absorption line indices as a function of both
stellar and stellar population parameters lags behind our understanding
of their redder counterparts.

This paper takes a closer look at the blue region using a time-tested
technique: the comparison of high-resolution model spectra with the
observations of a well studied {\it globular cluster} (GC). The first
test to which every population synthesis model must be submitted is
comparison of its predictions to the integrated spectra and colors of
GCs. We chose 47 Tucanae as the template for our study because it is among
the best studied Galactic GCs, being bright and nearby and located at
a high Galactic latitude where reddening is very low. It is also among
the most metal-rich GCs, with {\it [Fe/H]}=--0.7 (Carretta \& Gratton
1997), thus being a reasonably good match to the main characteristics
of elliptical galaxies, which are the ultimate targets of our study.

Reproducing the blue integrated spectrum of 47 Tuc is also interesting
in view of recent attempts to infer the cluster's spectroscopic age.
Gibson et al. (1999) employed stellar population synthesis models by Jones
\& Worthey (1995) to infer the spectroscopic age of the cluster on the
basis of the equivalent width of $H\gamma$. They found an exceedingly
high age of more than 20 Gyrs, which is in stark disagreement with the
values derived from CMD-fitting techniques (10-14 Gyrs, see Richer et
al. 1996, Salaris \& Weiss 1998, Liu \& Chaboyer 2000). Similar results
were found by Rose (1994) and Jones (1999).  More recently, Vazdekis
et al. (2001) showed that part of the problem is due to the fact that
the theoretical isochrones adopted in previous work did not take into
account important effects such as $\alpha$-enhancement and He-diffusion.
However, even with this correction a discrepancy of 4 Gyrs is still found
between their spectroscopic age and the one derived from CMD-fitting. By
looking at Figures 4 and 5 of Vazdekis et al., it can be seen that,
while the EW of $H\gamma$ is fitted with a 14 Gyr model, the isochrone
providing the best match to the cluster's CMD has an age of only 10 Gyrs.
It has even been suggested that the problem is more general, because
similar discrepancies were found by Cohen, Blakeslee \& Rhyzov (1998)
when comparing model predictions for $H\beta$ with observations of {\it
other} metal-rich Galactic GCs besides 47 Tuc.

In this and a companion paper (Schiavon et al. 2001, hereafter Paper
II), we take a new look at the discrepancy between spectroscopic and
CMD-based ages, on the basis of new models and a new approach. In this
first paper, the integrated spectrum of 47 Tuc is synthesized directly
from its color-magnitude diagram, without any input from theoretical
isochrones. Thus we are able to isolate and study effects which are due
purely to the characteristics of the spectral library. Hence real physical
problems in the evolutionary tracks and isochrones are separated from
limitations in the spectral library. A similar approach has been followed
before by de Souza, Barbuy \& dos Anjos (1993), Santos et al. (1995)
and Schiavon \& Barbuy (1999), for the synthesis of integrated spectra
of early-type galaxies and the red integrated spectrum of a metal-rich
GC in the Galactic Bulge, but it has never been applied before to the
blue integrated spectrum of 47 Tuc.

Another important issue is the effect of non-solar abundance ratios on
line indices. Specifically, empirical stellar libraries are generally
based on observations of stars from the solar neighborhood, whose detailed
abundance pattern may differ from that in other galaxies or even in
Galactic GCs. This is a matter of concern, since nearly all absorption
features in the integrated spectra of galaxies are due to a blend of
tens of lines from elements whose nucleosynthetic histories may be very
different. The problem is worse at blue wavelengths, where the crowding
of atomic and molecular lines is more severe than in the red. Moreover,
while the effect of non-solar abundance ratios on redder indices has
been estimated on the basis of spectrum synthesis from model photospheres
(Barbuy, Erdelyi-Mendes \& Milone 1992, Tripicco \& Bell 1995), no such
study has ever been performed on blue absorption line indices.

Again, 47 Tuc is a good template to be used in a study of abundance
ratio issues. It has been known for about three decades that the stars
in 47 Tuc and other GCs are characterized by a bimodal distribution in
abundance pattern: roughly half the stars are exceptionally N-enriched
while being mildly C-depleted, as opposed to the remaining stars,
with normal abundance ratios.  As a result, the stars with anomalous
abundance ratios have stronger CN-band strengths and weaker CH-band
strengths compared to the normal cluster stars, and also to field
stars of the same $T_{eff}$, luminosity and mass. There still is no
unanimous explanation for this phenomenon, especially in view of the
fact that the bimodality extends down to stars below the turn-off (see
Cannon et al. 1998; Cohen, Behr \& Briley 2001 and references therein),
so that mixing cannot be responsible for the effect.  Anomalous CNO
abundance ratios is a very important issue for population synthesis in
the blue, given that the spectral restframe interval $\lambda\lambda$
3800-4500 {\AA} is plagued by thousands of lines due to CN and CH.
Indeed these lines are found in close proximity to the $H\delta$ and
$H\gamma$ indices, and it is possible that they might have disturbed
previous spectroscopic ages, an issue we will examine in some detail.
In this paper, we study the abundance anomaly from a spectrum synthesis
standpoint, by computing stellar synthetic spectra with different
abundance ratios, based on modern model atmospheres and comprehensive
atomic and molecular line lists (Castilho 1999). On the basis of such
computations, we estimate the effect of C,N abundance variations on key
line indices.  The corrections derived in this paper are then applied in
Paper II to computations of model spectra based on theoretical isochrones,
based upon which we determine the spectroscopic age of 47 Tuc.

The paper is laid out as follows: in Section 2, we describe the spectral
library employed in our models and briefly present the determination of
the fundamental parameters of the library stars.  In Section 3, we present
a concise discussion of the behavior of important spectral indices as
a function of stellar parameters. In Section 4 the procedure to compute
integrated model spectra is described. In Section 5 model predictions are
compared to the observations of 47 Tuc. Our conclusions are summarized in
Section 6. In Appendix A we describe our synthetic spectrum computations
and study the effect of C,N abundance variations on blue line indices.

\section{Spectral Library} \label{speclib}

The adequacy of a spectral library for population synthesis in the blue
depends importantly on spectral resolution. High resolution is needed to
disentangle the contribution of different lines to the broad features
observed in $\sigma$-smoothed integrated spectra of galaxies. This is
specially relevant in the blue spectral region, which is characterized
by the crowding of hundreds of thousands of lines. The spectral library
that best meets our needs is the one by Jones (1999). It is relatively
comprehensive, containing 624 stars with spectral types ranging between
O and M, all luminosity classes and metallicities within the interval
-2.5$\leq{\it [Fe/H]}\leq$+0.3. The resolution is moderately high (1.8 {
\AA}, FWHM), and it is flux calibrated and has high S/N.  A more detailed
description can be found in Jones (1999) and Vazdekis (1999). The spectra
were collected by Lewis Jones using the Coud\'e-feed spectrograph at
KPNO in two wavelength intervals: $\lambda\lambda$3820--4510 {\AA}
and $\lambda\lambda$4780--5470 {\AA}.

\subsection{Atmospheric Parameters of Library Stars} \label{stelpar}

In what follows, we provide a description of the determination of the
atmospheric parameters of the stars from the library.  A more complete
discussion of our method and the full list of stellar parameters will
be presented elsewhere (Schiavon 2002, in preparation).

The fundamental stellar parameters we need to determine for each library
star are $T_{eff}$, metallicity and surface gravity ($\log g$). The
highest possible accuracy is required for $T_{eff}$ and metallicity. The
accuracy requirement on $\log g$ is lower because of the relatively
lower influence of this parameter on stellar spectra.

Most of the stars from the Jones library have been the subject of
detailed abundance analyses in the past, and their atmospheric parameters
can be found in the Cayrel de Strobel et al. (1997) catalog of {\it [Fe/H]}
determinations. However, the stellar parameters listed in this catalog
are very heterogeneous. They were determined using a variety of methods,
based on data of varying quality and employing different sets of model
atmospheres. As a consequence, the different sets of stellar parameters
are often in significant disagreement (see discussion in Soubiran, Katz \&
Cayrel 1998). Having a homogeneous set of stellar parameters is of great
importance for our purposes, so we decided to determine our own set of
stellar parameters for all stars in the library in a consistent fashion.

Besides homogeneity, absence of any important systematic effect in the
stellar parameters is another key requirement. The best way to check for
such effects is by comparing our stellar parameters with other sets and
also by verifying that they are consistent with empirical color $vs.$
($T_{eff}$,{\it [Fe/H]}) calibrations from the literature. Such checks are
also important for deciding which is the most appropriate and consistent
$(B-V)$ vs. $T_{eff}$ calibration to adopt when transforming 47 Tuc's
observed CMD to the ($T_{eff}$,$\log g$) plane.

The determination of the atmospheric parameters of the library stars
followed different procedures for giants and dwarfs. For dwarfs, the
parameters were derived from Str\"omgren photometry together with
up-to-date calibrations from the literature.  For giants, they are
based on spectral line indices, whose dependence on stellar parameters
were determined by ourselves. The procedures are described in the next
two subsections.  The stellar colors were dereddened as follows:
for roughly one third of the library stars, $E(B-V)$ values were taken
from Neckel et al. (1980), Savage et al. (1985), Carney et al. (1994),
Blackwell \& Linas-Gray (1994), Alonso, Arribas \& Mar\'\i nez-Roger
(1996) and Beers et al. (1999). For the stars lacking previous reddening
determinations, these values were determined from their Galactic
coordinates and parallaxes by adopting the extinction model by Arenou,
Grenon \& G\'omez (1992), as implemented in a FORTRAN code by Hakkila
et al. (1997). Arenou et al.'s extinction model was obtained by fitting
smooth functions to extinctions and distances of over 40000 stars.
The agreement between $E(B-V)$s so determined with the values collected
from the literature is good, within 0.03 mag. Adopting another
model for interstelar extinction, like the one by Chen et al. (1999),
which incorporates the more recent, higher-resolution extinction maps
by Schlegel, Finkbeiner \& Davis (1998), results in systematic, but
minor changes in the final stellar parameters of dwarf stars ($\simless$
50K in Teff and $\simless$ 0.03 dex in [Fe/H]). As we show in Paper II,
this has a minor effect on model predictions (see Table 2 of Paper II).
A more detailed discussion is deferred to a future paper (Schiavon 2002,
in preparation). We also compiled $(B-V)$s for all the library stars
from the SIMBAD database, correcting them for reddening in the same way.

\subsubsection{Dwarfs} \label{dwarfs}

Metallicities and $T_{eff}$s for the dwarfs were determined from
Str\"omgren photometry. We chose to adopt Str\"omgren photometry because
of the availability of such data for most of the library dwarfs, and
also because reliable calibrations of Str\"omgren indices as functions
of $T_{eff}$ and {\it [Fe/H]} exist in the literature.

Effective temperatures and metallicities of the dwarf stars were
estimated from their Str\"omgren $(b-y)$, $\beta$, $c_1$ and $m_1$
indices.  For cool stars ($\beta < 2.7$), we adopted the color vs.
temperature calibrations of Alonso, et al. (1996) and Schuster \& Nissen
(1989). For hotter stars, we adopted the calibrations of Smalley \&
Dworetsky (1993) and the standard sequence from Perry, Olsen \& Crawford
(1987). Str\"omgren photometry was available for the vast majority (90\%)
of our sample in the catalog of Hauck and Mermilliod (1998). These stars
were used to build relations between photometric indices and spectral
features selected according to their sensitivity to a given stellar
parameter. The relations were then used to estimate the Str\"omgren
indices of the remaining library stars. Surface gravities were inferred
from the absolute magnitudes (all relying on Hipparcos parallaxes) and
$T_{eff}$s in the classical way, adopting bolometric corrections derived
from stellar $T_{eff}$s and {\it [Fe/H]}s through the calibrations of
Alonso, Arribas \& Mar\'\i nez-Roger (1995). Masses were interpolated
in the 5 Gyr isochrones of Girardi et al. (2000) for a given $T_{eff}$
and metallicity.

The above procedure was applied to all dwarfs inside the range 4500 $\leq$
$T_{eff}$ $\leq$ 7000 K, which is the interval inside which the above
calibrations are valid. For cooler and hotter stars, we selected
the most recent stellar parameter determinations listed in the catalog
of Cayrel de Strobel et al. (1997).

\begin{figure}[htpb]
\plotone{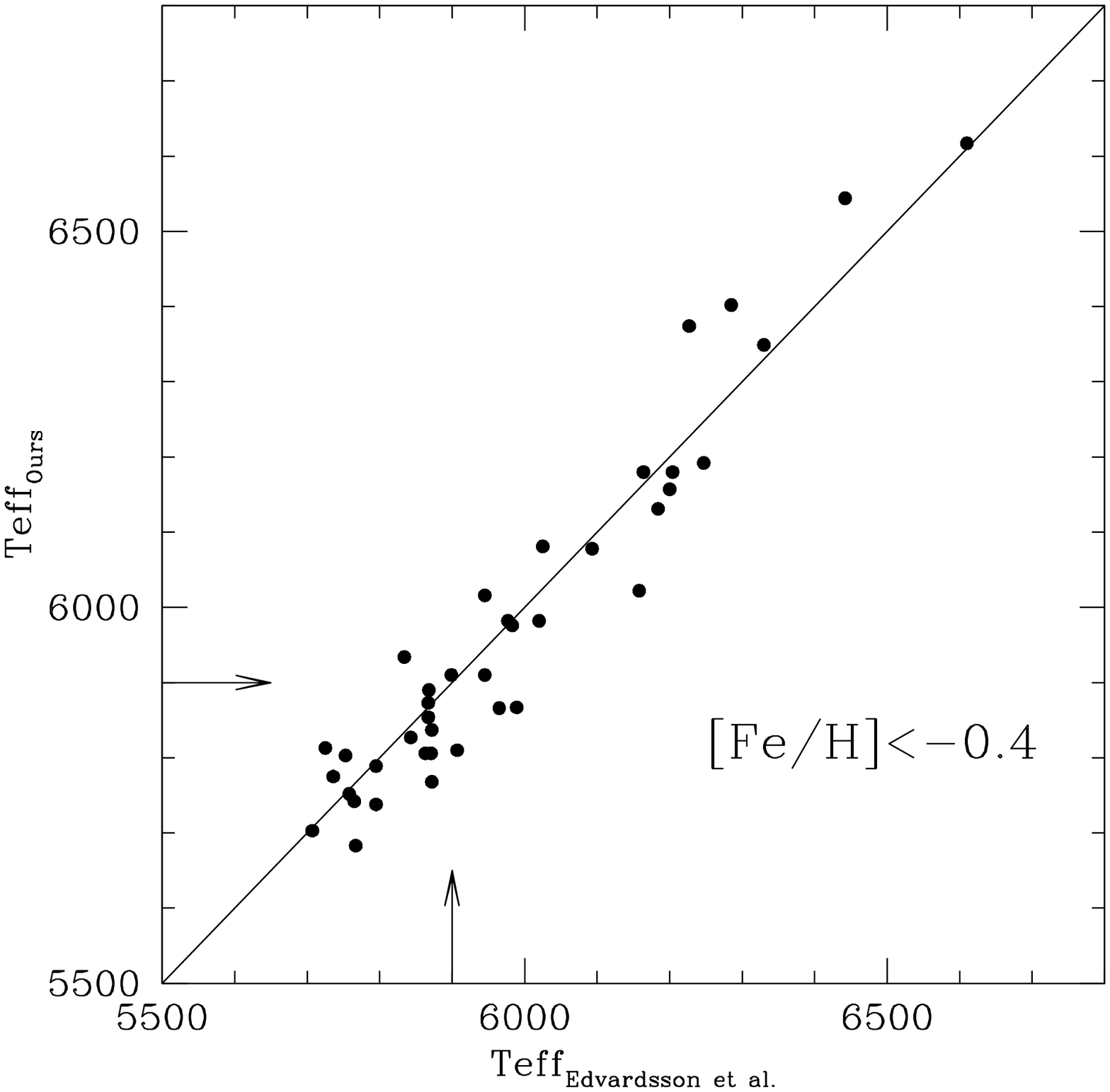}
\caption{Comparison between our $T_{eff}$s and those from Edvardsson et
al. (1993) for metal-poor dwarfs. There are slight differences between
the two data sets: our $T_{eff}$s are lower by $\sim$ 50 K for stars
cooler than 6200 K and higher by $\sim$ 70 K for hotter stars. The arrows
indicate the $T_{eff}$ of turn off stars in 47 Tuc.
\protect\label{fig1a}}
\end{figure}

As a check on our $T_{eff}$ and metallicity scales, we compare our values
to determinations by Edvardsson et al. (1993) for stars in common with
our library. The comparison between the two sets of $T_{eff}$s is shown
in Figure~\ref{fig1a}. In this figure, the approximate $T_{eff}$ of
the turn-off stars in 47 Tuc is indicated. For clarity, the comparison
between the two sets of $T_{eff}$s is restricted to stars with {\it
[Fe/H]}$<$--0.4, which are the ones that enter the synthesis of the
integrated spectrum of 47 Tuc.  We discuss the determination of stellar
parameters of more metal-rich stars in our forthcoming paper (Schiavon
2002, in preparation). The agreement between the two sets of $T_{eff}$s is
generally good. However, there are slight disagreements: for stars cooler
than 6200 K, our $T_{ eff}$s are lower by as much as $\sim$35 K, and for
hotter stars they may be hotter by $\sim$60 K. The latter mismatch is
hard to assess since there are only 5 stars in common with Edvardsson et
al. with $T_{eff} > $6200 K. It is also less important for the purposes
of this paper, given that the turn-off of 47 Tuc is located at lower
$T_{eff}$s, as indicated in Figure~\ref{fig1a}.  It is also important
to note that no such mismatch is seen when the above comparison is done
for more metal-rich stars. It is difficult to assess the reason for the
differences seen for the more metal-poor stars. Like ours, Edvardsson
et al.'s $T_{eff}$s are inferred from Str\"omgen photometry. Since
our determinations rely on the latest calibrations, we do not correct
our values. In fact, the effect of these uncertainties in the $T_{eff}$
and metallicity scales of dwarf stars on our predicted line strengths is
very small as compared with other more important effects (see discussion
in Section \ref{unc} and in Paper II).

\begin{figure}[htpb]
\plotone{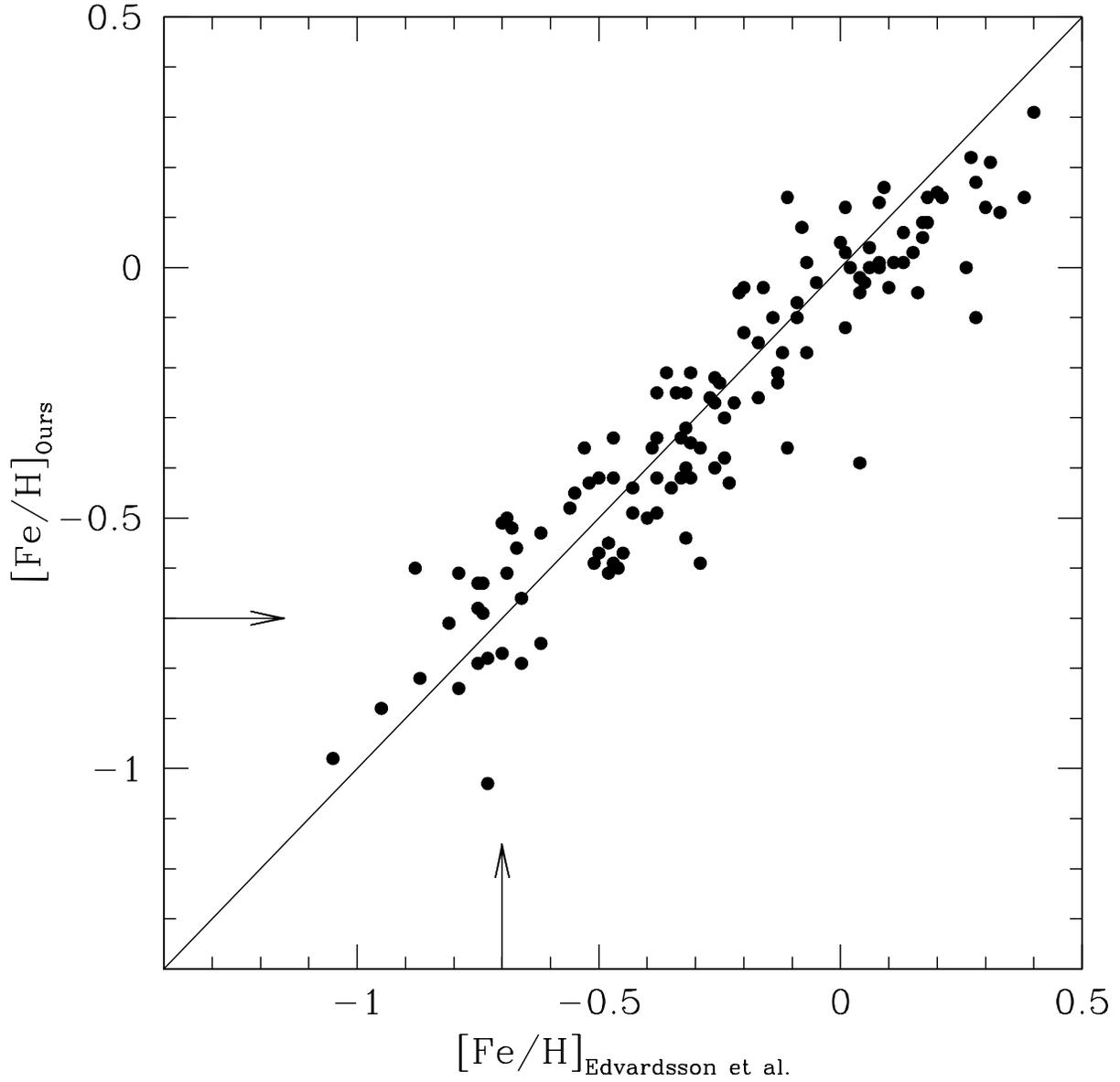}
\caption{Comparison between our {\it [Fe/H]}s for library dwarfs and those
from Edvardsson et al. (1993). They are in good agreement for stars more
metal-poor than {\it [Fe/H]}=0. The arrows indicate the nominal metallicity
of 47 Tuc.  
\protect\label{fig1b} } 
\end{figure}

In Figure \ref{fig1b} we compare our values for {\it [Fe/H]} with
those from Edvardsson et al. Except for the most metal-rich stars ({\it
[Fe/H]}$>$0), for which our values are on average lower by $\sim$ 0.1 dex,
the two datasets compare very well. Most importantly for the purposes of
this paper, the agreement is very good around the metallicity of 47 Tuc
($[Fe/H]\sim-0.7$). This is reassuring, given that the two sets of {\it
[Fe/H]}s are reasonably independent, Edvardsson et al.'s values being
based on classical detailed abundance analyses of high resolution spectra.

\begin{figure}[htpb]
\plotone{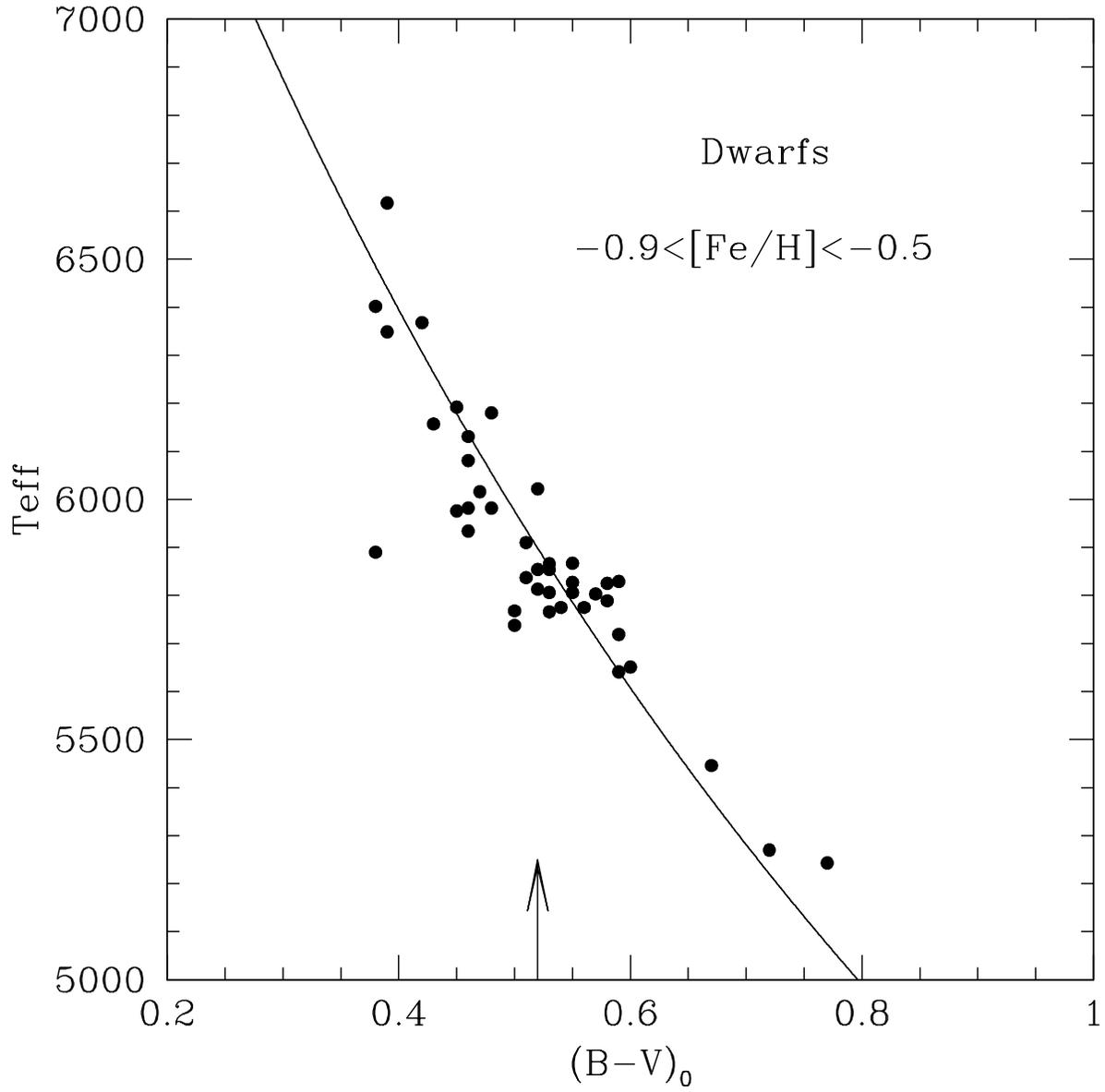}
\caption{Comparison of the $(B-V)_0$ $vs.$ $T_{eff}$ relation for
our metal-poor dwarfs with the calibration from Alonso et al. (1996),
computed for {\it [Fe/H]}=--0.7. The arrows indicate the approximate
position of the dereddened turn-off of 47 Tuc, where the agreement is
very good.
\protect\label{fig1c}
}
\end{figure}

As a further check on our parameters, we display in Figure \ref{fig1c}
the metal-poor dwarfs from the library in the $(B-V)_0$ $vs.$ $T_{eff}$
plane, overlaid by the calibration from Alonso et al. (1996) for {\it
[Fe/H]}=--0.7. From this figure, it can be seen that our $T_{eff}$s are
in good agreement with the calibration from Alonso et al. except for K
dwarfs with $(B-V)>$0.65, where our $T_{eff}$s seem to be slightly too
high ($\sim$100 K). Unfortunately, there are too few K dwarfs in our
library to allow for a more reliable assessment of this effect. Again,
for our purposes in this work, it is important to notice that in the
region of the turn-off of 47 Tuc (indicated in the figure), the agreement
between our data and Alonso et al.'s calibration is satisfactory. The
discrepancy in the K dwarf regime is of minor importance because these
stars are responsible for a very small contribution to the blue integrated
light of the cluster (see discussion in Section  \ref{contrib}).

Figure \ref{fig1c} is of particular relevance for yet another reason:
in our procedure to compute the integrated spectrum from the CMD of the
cluster, we first transform the colors and absolute magnitudes coming from
the CMD into $T_{eff}$ and $\log g$, on the basis of which the library
stars that enter the synthesis are selected. Therefore, consistency
demands that we adopt a calibration of $(B-V)_0$ $vs.$ $T_{eff}$ for
47 Tuc that is in agreement with the $T_{eff}$s and dereddened $(B-V)$s
of the library stars. The agreement seen in Figure \ref{fig1c} suggests
that, at least for the dwarfs, we can adopt Alonso et al.'s calibration
for 47 Tuc without incurring any important inconsistency.

\subsubsection{Giants} \label{giants}

A different approach had to be followed to determine the stellar
parameters of our library giants. Although accurate temperatures can be
inferred from reliable calibrations of metallicity-insensitive colors such
as $(V-I)$ (Alonso Arribas \& Mart\'\i nez-Roger 1999), it is not possible
to do the same for metallicities of giant stars. For this reason, our
$T_{eff}$s and metallicities for giants rely on measurements of absorption
features in the spectral library and calibrations of those features
against stellar parameters, built specially for our purpose. Surface
gravities are estimated from isochrones, using an assumed mass and an
estimated $T_{eff}$, absolute magnitude and bolometric correction.

The starting point of our method is to select stars in the library with
the most reliable stellar parameter determinations from the literature.
For that purpose, we culled stars from the library in common with
the works of Alonso et al. (1999), Soubiran et al. (1998), Luck \&
Challener (1995) and McWilliam (1990). From Alonso et al. we took only
$T_{eff}$s, since their {\it [Fe/H]}s come from Cayrel de Strobel et al.'s
catalog. Using stars in common to these four data sets, we adjusted all
$T_{ eff}$s and {\it [Fe/H]}s to the scales of Soubiran et al., because
they are based on a very comprehensive data set and have the most stars
in common with our library. The stellar parameters in the Soubiran et
al.'s work are based on determinations collected from the literature,
among which those that best match their observed high-resolution spectra
are selected. The average offsets applied were of the order of $\sim$70 K
in $T_{eff}$ and 0.1 dex in {\it [Fe/H]}.  With this procedure, we were
able to gather a set of 254 ``template'' giants with $T_{eff}$ and {\it
[Fe/H]} on the scales of Soubiran et al. These were the basis on which
a calibration of spectral features against stellar parameters was built.

The next step was to select those spectral features that are the best
indicators of $T_{eff}$ and {\it [Fe/H]}. By studying the behavior of
a set of Balmer and Fe line equivalent widths and line depth ratios as
a function of those parameters, we finally selected $H\beta$ and the
average of the indices Fe4383, Fe5270 and Fe5335 (henceforth $Fe_m$)
as the best pair to determine $T_{eff}$ and {\it [Fe/H]}. $H\beta$
is predominantly sensitive to $T_{eff}$, but it is also influenced by
metallicity, being stronger in the spectra of more metal rich giants
(cf. Figure \ref{fig3c} below). The dependence on metallicity is due to
contamination by metal lines in the index passband. On the other hand,
$Fe_m$ is strongly sensitive to both $T_{eff}$ and {\it [Fe/H]}.

The pseudocontinua and bandpasses used for $H\beta$, Fe4383, Fe5270
and Fe5335 were those given by Worthey et al. (1994). The indices were
measured in the spectra of the template giants from the library, after
convolving the spectra to a velocity dispersion of $\sigma\sim 100$ km
s$^{-1}$, in order to bring them to the same system as the observations
of 47 Tuc used in this work. It is important to stress that, even though
the passband definitions of Worthey et al. were adopted, our measurements
are {\it not} on the Lick-IDS system since the spectra are not convolved
to the much lower resolution of Lick-IDS spectra. In this way, we take
full advantage of the higher resolution of our library spectra.

We performed a one-error least-squares regression to the functions
{\it [Fe/H]}($Fe_m$,$H\beta$) and $T_{eff}$ ($H\beta$,$Fe_m$,{\it [Fe/H]}).
Satisfactory fits were found to polynomials of the form

\begin{equation}
{\it [Fe/H]}\,\,=\,\,a_0\,\,+\,\,a_1\,Fe_m\,\,+\,\,a_2\,Fe_m^2\,\,+\,\,a_3\,Fe_m\,
H\beta\,\,+\,\,a_4\,H\beta\,\,+\,\,a_5\,H\beta^2,
\end{equation}
and
\begin{equation}
T_{eff}\,\,=\,\,b_0\,\,+\,\,b_1\,Fe_m\,\,+\,\,b_2\,Fe_m^2\,\,+\,\,
b_3\,Fe_m\,H\beta\,\,+\,\,b_4\,H\beta\,\,+\,\,b_5\,H\beta^2\,\,+\,\,
b_6\,Fe_m\,{\it [Fe/H]}.
\end{equation}
The coefficients of the fits are listed in Table~\ref{tbl-1}.
One-$\sigma$ errors in $T_{eff}$ and {\it [Fe/H]} are $\pm$ 80 K and
$\pm$ 0.1 dex respectively.

Surface gravities of the giants were determined in the same way as for
the dwarfs, inferring absolute magnitudes from Hipparcos parallaxes,
adopting 1 M$_\odot$ for all stars, and computing bolometric corrections
from the calibration by Alonso et al. (1999).

\begin{figure}[htpb]
\epsscale{0.8}
\plotone{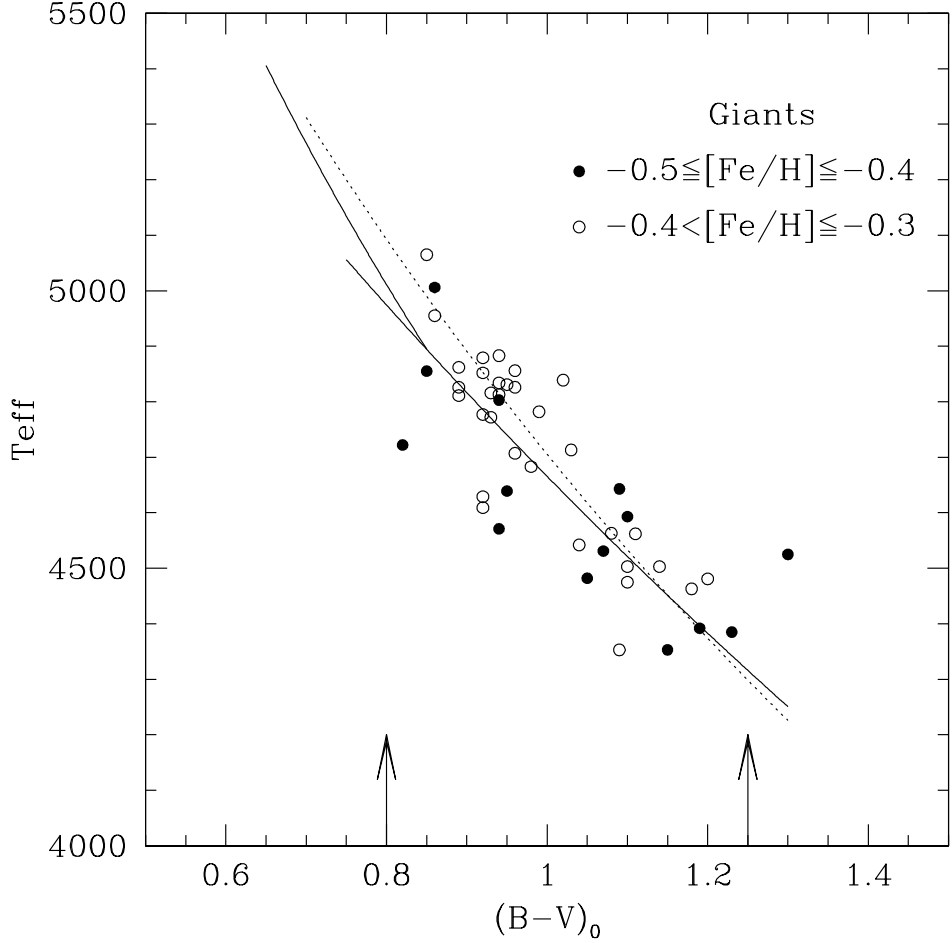}
\caption{A comparison between $T_{eff}$ and $(B-V)_0$ for the metal-poor
giants of the library versus the calibration from Alonso et al. (1999) for
{\it [Fe/H]}=--0.4 (dotted line). The solid lines correspond to our adopted
calibration. The curve displayed for $T_{eff}$s lower than $\sim$ 5000
K corresponds to our fit to the data (eq. 3) and the one for higher
$T_{eff}$s is the Alonso et al. (1999) calibration for stars bluer than
$(B-V)$=0.8 displaced by --100 K in order to guarantee a smooth transition
to our calibration. Stars within the $(B-V)$ interval bracketed by the
values indicated by the arrows contribute 80\% of the giant light at
$\sim$ 4500 {\AA} 
\protect\label{fig2a} 
}
\end{figure}

By construction, our $T_{eff}$ and metallicity scales for giants are
consistent with those of Soubiran et al. (1998). Thus we provide here only
a comparison of our values for the metal-poor giants with the $(B-V)_0$
$vs.$ $T_{eff}$ calibration by Alonso et al. (1999), keeping in mind that
differences between Soubiran et al. and the works of Luck \& Challener
(1995) and McWilliam (1990) amount to up to 70 K and 0.1 dex in $T_{eff}$
and {\it [Fe/H]} respectively. This is shown in Figure \ref{fig2a}, where
we compare library giants of $[Fe/H]\sim$--0.4 with Alonso et al.'s
calibration for {\it [Fe/H]}=--0.4 (solid line). The lack of a sufficient
number of giants in the library with $[Fe/H]\simless$--0.5 prevents us
from making this test for lower metallicities. From this figure, it can
be seen that a reasonably good agreement is achieved. However, there
is a slight discrepancy in the sense that Alonso et al.'s calibration
seems to provide $T_{eff}$s slightly hotter ($\sim$ 50--100 K) than our
values for stars bluer than $(B-V)$ $\sim$ 1.

It is difficult to figure out the source of the differences apparent
in Figure  \ref{fig2a}. As mentioned before, the procedure we follow
to synthesize the integrated spectrum of 47 Tuc from the CMD involves
converting colors and absolute magnitudes into $T_{eff}$s and $\log g$s.
For this purpose, a $T_{eff}$ vs. $(B-V)$ transformation is necessary
for the dwarfs and giants separately.  It is clear that we would be
introducing systematic effects in our computations if we adopted Alonso
et al.'s calibration for the giants, because it is inconsistent with
the $T_{eff}$ $vs.$ $(B-V)$ relation built into our spectral library.
We therefore derived our own $T_{eff}$ $vs.$ ($(B-V)_0$,{\it [Fe/H]})
relation, based on our $T_{eff}$ and {\it [Fe/H]}-scales and the dereddened
colors of the library giants. By using this new calibration, the derived
$T_{eff}$s and $\log g$s for 47 Tuc stars will be fully consistent with
the library $T_{eff}$ and {\it [Fe/H]}-scales derived in this section. Thus,
we fitted to our data a polynomial of the form

\begin{equation}
\theta_{eff} = a_0\,\,+\,\,a_1\,(B-V)\,\,+\,\,a_2\,(B-V)^2\,\,+\,\,a_3\,
(B-V)\,{\it [Fe/H]}\,\,+\,\,a_4\,{\it [Fe/H]}\,\,+\,\,a_5\,{\it
[Fe/H]}^2, 
\end{equation}
where $\theta_{eff}$=5040/$T_{eff}$.  The coefficients of this polynomial
fit are shown in Table~\ref{tbl-2a}, and the interval of its applicability
is 0.8$<(B-V)<$1.2, --0.7$\leq[Fe/H]\leq$+0.2. The 1-$\sigma$ error in
$T_{eff}$ is 66 K. The polynomial is overlaid on the data in Figure
\ref{fig2a} (dotted line), where it can be seen that adoption of our
calibration versus Alonso et al.'s may lead to a difference of up to
100 K at $(B-V)$=0.8. The effect of uncertainties in the $T_{eff}$ and
metallicity-scales on our predictions is discussed in Section~\ref{unc}
and Paper II. The paucity of hot metal-poor giants in our library limits
the applicability of this relationship to $T_{eff}>$5000 K. This is not
very important because the vast majority of the giants in 47 Tuc are
cooler than 5000 K. For the few HB stars with $T_{eff} > $ 5000 K, we
adopted Alonso et al.'s scale, for the corresponding $(B-V)$ interval,
with a zero-point correction of 100 K in order to guarantee a smooth
transition between the two calibrations around $(B-V)\sim$0.8. This is
shown as the short solid line in Figure \ref{fig2a}.

\subsubsection{Final parameters} \label{finpar}

The last step in our procedure was to plot the Balmer line strengths
of all stars (dwarfs and giants separately) versus $T_{eff}$ for
different metallicities, removing from the library all stars that were
very deviant from the general trends.  In this way we made sure that
our predictions of Balmer line intensities are free of the influence
of stars with very unusual Balmer line equivalent widths for their
atmospheric parameters. The final pruned library is comprised of 564
stars: 353 giants and 211 dwarfs.

\begin{figure}[htpb]
\plotone{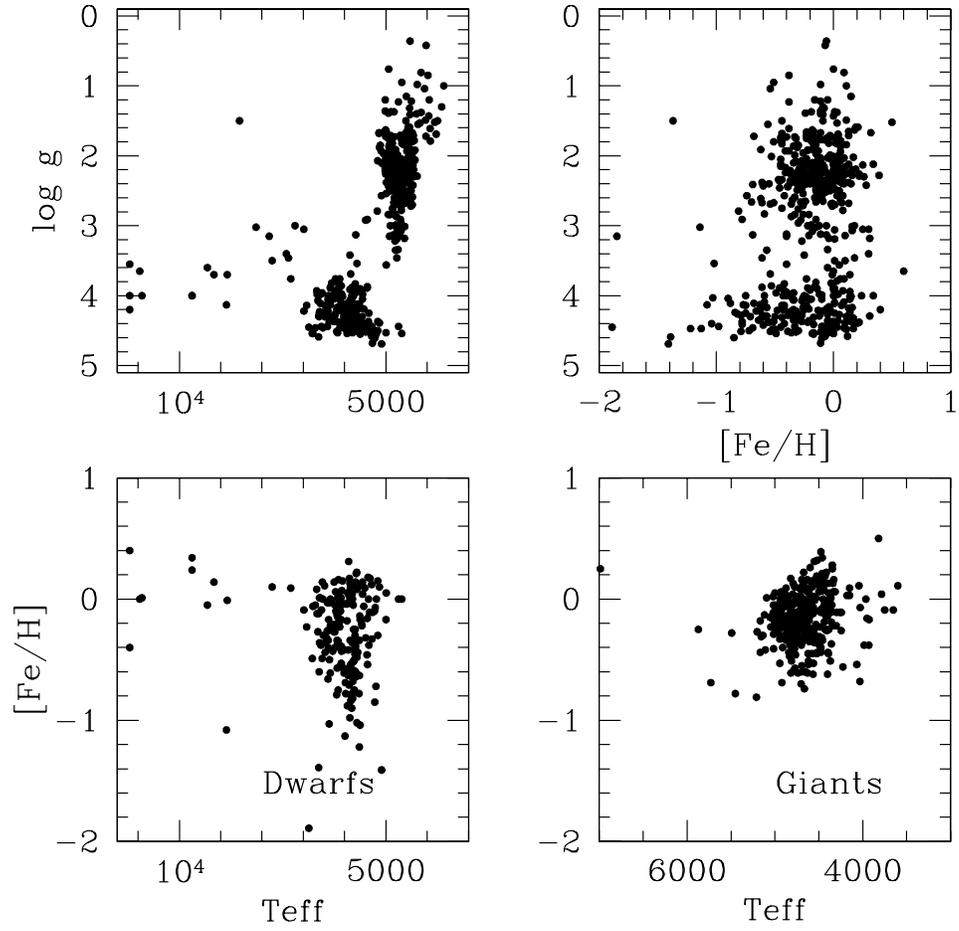}
\caption{Distribution of the library stars in stellar parameter space.
\protect\label{fig3}
}
\end{figure}

Figure \ref{fig3} shows the distribution of the finalized library stars
in the ($T_{eff}$, {\it [Fe/H]}), ($T_{eff}$, $\log g$) and ($\log
g$, {\it [Fe/H]}) planes. From the plots, it can be seen that the
spectral library covers large portions of the stellar parameter space
well. However, even though this is among the largest of the empirical
spectral libraries so far employed in stellar population synthesis, it
has important limitations. In particular, the library is sparse in two
regions: for dwarfs hotter than $\sim$7000 K and for metal-poor giants
($[Fe/H]\simless$--0.5). The first limitation may introduce artificial
discontinuities in the behavior of line indices as a function of age,
but it only affects models of stellar populations younger than 4 Gyrs
(at solar metallicity) and therefore is not important for the analysis
carried out in this paper.  The second limitation is of great concern
for this paper, since the metallicity of 47 Tuc ({\it [Fe/H]}=--0.7)
is within the range where the spectral library is not very dense,
particularly for giants. We come back to this issue below.

\section{Line indices as a Function of Stellar Parameters} \label{indstar}

Before starting our discussion of the integrated spectrum of 47 Tuc, it
is of interest to first review the dependence of absorption line indices
of individual stars as a function of stellar parameters, as this will
help us to better understand the behavior of our models as a function
of model inputs. Definitions of the line indices studied can be found in
Worthey et al. (1994) in most cases, except for $H\delta_F$, defined by
Worthey \& Ottaviani (1997), $H\gamma_{HR}$, defined by Worthey \& Jones
(1995) and $H\gamma_{\sigma < 130}$, defined by Vazdekis et al. (2001).

Even though the high-resolution of both the observed spectrum of 47
Tuc and the spectral library used in our model construction allows us to
explore narrow-band indices like Ca4227$_{HR}$ and $H\gamma_{HR}$ (Worthey
\& Jones 1995), we will refrain from adopting them.  These indices have
extremely narrow passband and pseudocontinuum windows, which makes them
especially sensitive to errors in $gf$-values, when computing synthetic
stellar spectra. Because we adopt spectrum synthesis in order to correct
our model predictions for the effect of abnormal C,N abundances in 47
Tuc stars (see Section~\ref{synth} and the Appendix), we concentrate
our discussion for the rest of this paper on wider-band indices like
$H\gamma_{\sigma<130}$ and Lick Ca4227, for which such corrections
are far more robust.  Moreover, as the ultimate goal of this work is
to calibrate models to be applied to the analysis of $\sigma$-smoothed
spectra of elliptical galaxies, it is more sensible that we concentrate
our efforts on wide-band indices.

The size and breadth of our spectral library make it easy to gauge the
effect of $T_{eff}$ and metallicity on line index strength. The same
does not apply to $\log g$, however, because library dwarfs and giants
are in common only within a very narrow $T_{eff}$ interval. Therefore,
all comments below about the dependence of line indices on $\log g$
are restricted to the interval 4600 $\simless T_{eff} \simless$ 5500 K.

\begin{figure}[htpb]
\plotone{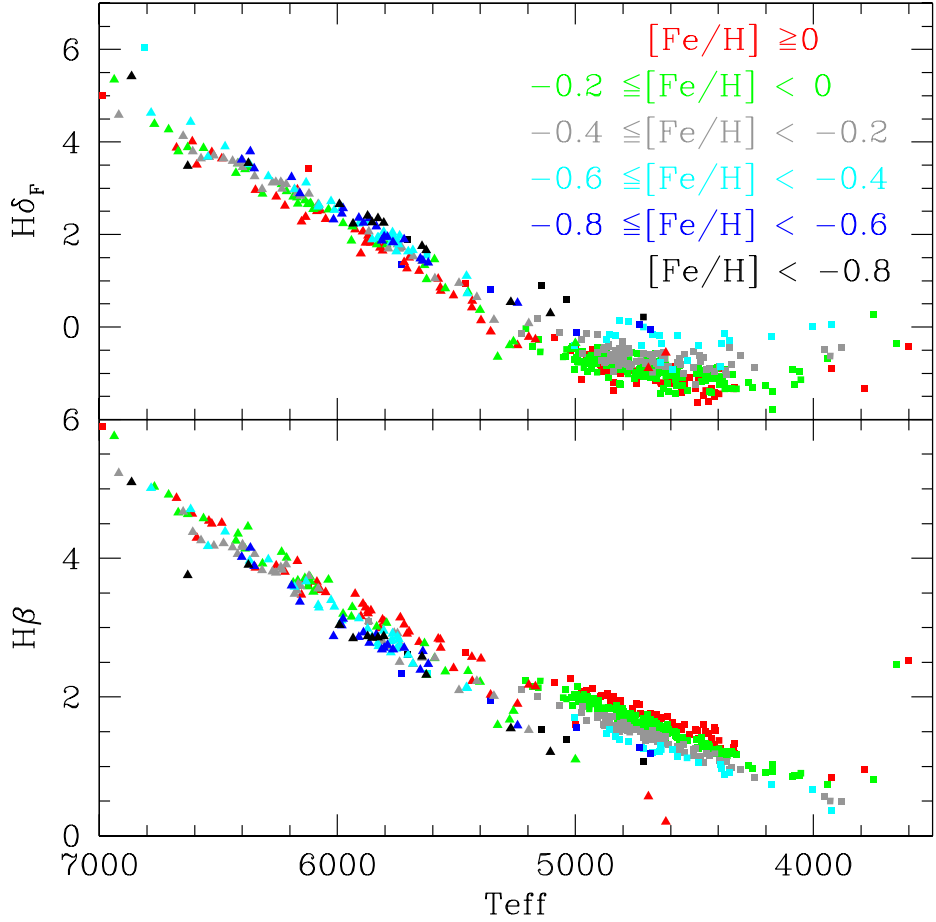}
\caption{$H\beta$ and $H\delta_F$ as a function of $T_{eff}$ and
metallicity. Dwarfs are represented as triangles, and giants as
circles. The effect of $\log g$ can be seen in the range 5000--5700  K,
where giants and dwarfs overlap. Color codes for metallicity are shown
in the upper right corner of the upper panel. $T_{eff}$ dominates over
both metallicity and gravity. The gravity discontinuity at $T_{eff}$
$\sim$ 5200 K is small. Metallicity effects are evident, especially in
the giants, but they have opposite sign in the two indices, $H\delta_F$
weakening with metallicity and $H\beta$ strengthening.
\protect\label{fig3c} } 
\end{figure}
\begin{figure}[htpb]
\plotone{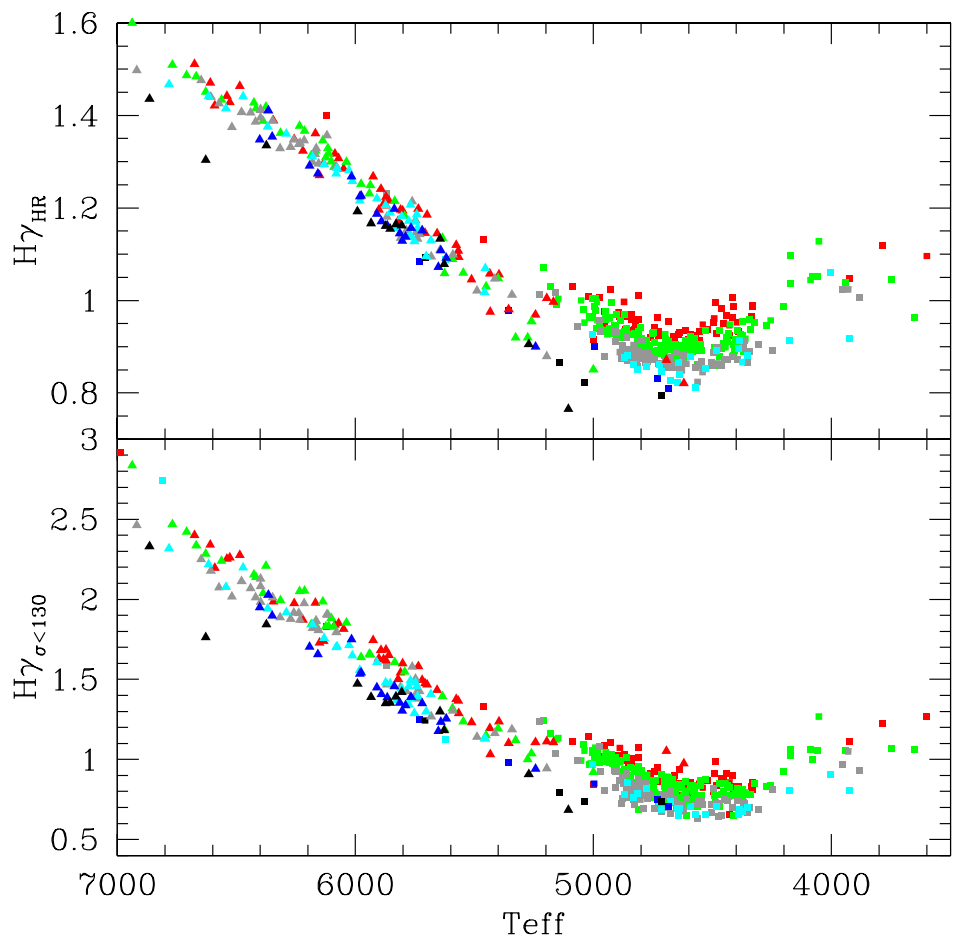}
\caption{The same as Figure \protect\ref{fig3c} but for $H\gamma_{HR}$ and
$H\gamma_{\sigma < 130}$. These two versions of $H\gamma$ are similar but
subtly different. To first order, they resemble $H\beta$ and $H\delta_F$
in Figure \protect\ref{fig3c}. However, their metallicity sensitivities are
somewhat larger, both, like $H\beta$, strengthening with {\it [Fe/H]}. Their
total throw is also considerably smaller, despite the fact that the
intrinsic equivalent width of $H\gamma$ is similar to $H\beta$ and
$H\delta_F$.  See discussion in the text.
\protect\label{fig3d} }
\end{figure}
\begin{figure}[htpb]
\plotone{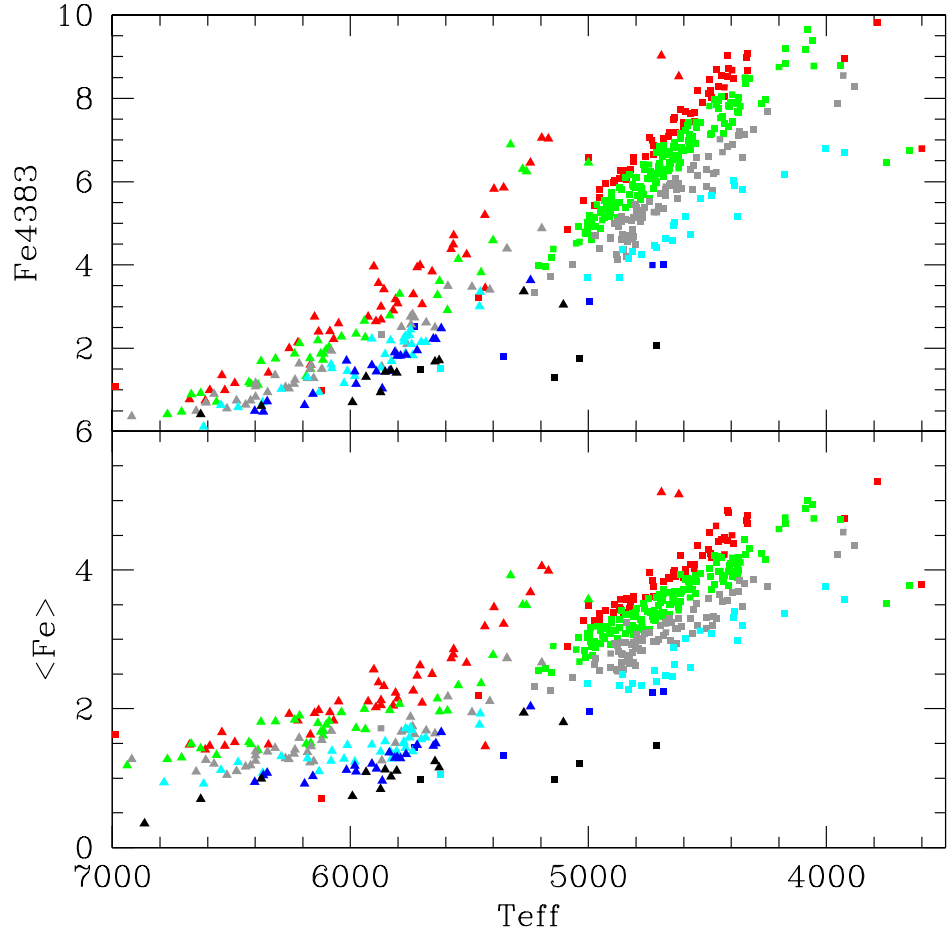} \caption{The same as Figure
\protect\ref{fig3c} but for  $<Fe>$ and Fe4383. $T_{eff}$ is the prime parameter
influencing index strength, followed by [$Fe/H$] in the case of these two
indices. Some gravity effects are seen near $T_{eff}$ $\sim$ 5200 K.
\protect\label{fig3a} } 
\end{figure}
\begin{figure}[htpb]
\plotone{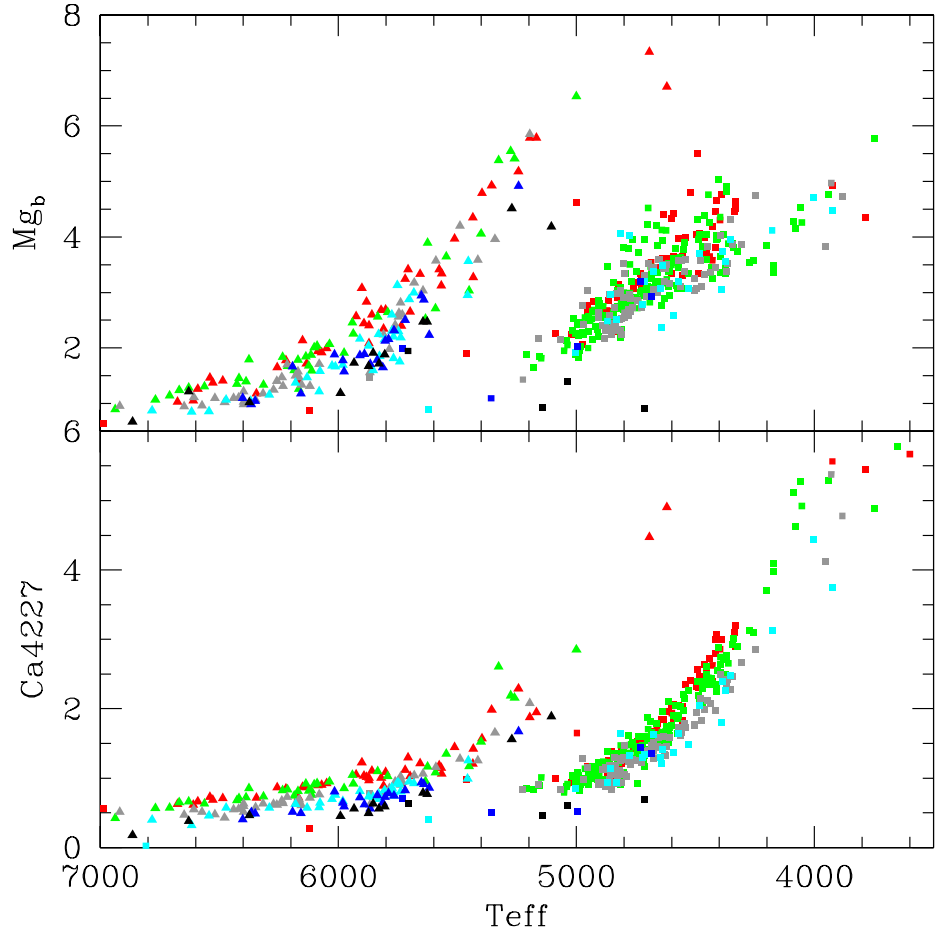}
\caption{The same as Figure \protect\ref{fig3c} but for $Mg\,b$ and Ca4227. 
Gravity now eclipses metallicity, as shown by the strong discontinuity
between dwarfs and giants at $T_{eff}$ $\sim$ 5200 K.
\protect\label{fig3b}
}
\end{figure}

The indices of most importance to us are plotted against $T_{eff}$ in
Figures \ref{fig3c} to \ref{fig3b}. Dwarfs are represented by triangles
and giants by circles.  Metallicities are color-coded, as shown in the
upper panel of Figure \ref{fig3c}. For all indices, $T_{eff}$ is the
prime parameter influencing index strength.  The second most important
parameter is {\it [Fe/H]} in the case of $<Fe>$ and Fe4383, and $\log
g$ in the case of $Mg\,b$ and Ca4227.  For $H\beta$ and $H\gamma$, {\it
[Fe/H]} and $\log g$ seem to have approximately the same impact on line
strength. $H\delta_F$ appears to be the only index which is virtually {\it
insensitive} to $\log g$.  We comment below on the features particular
to each of these groups.

\noindent {\it Balmer lines}: all Balmer line indices are strongly
sensitive to $T_{eff}$, being stronger in the spectra of hotter stars
because of increasing excitation. They are also mildly sensitive to
$\log g$, being stronger in the spectra of giants at cool temperatures.
This is because of the decrease of continuum opacity (mostly due to H$^-$)
as electron pressure decreases. Below 4000 K, $H\beta$ increases for
lower $T_{eff}$s because of the presence of a strong TiO bandhead ($\sim$
4851 {\AA}) within the passband of the index.  $H\gamma$ and $H\delta$
undergo similar (but less drastic) upturns below $T_{eff}$ $\sim$ 4500
K, because at low $T_{eff}$s the indices are dominated by contaminating
metal lines, which get stronger for decreasing $T_{eff}$. All Balmer
line indices are sensitive to {\it [Fe/H]}. Both $H\beta$ and $H\gamma$
become stronger for higher {\it [Fe/H]} because of the presence of metal
lines in the index passband. The opposite trend is seen in $H\delta$,
for which the contamination is more important in the pseudocontinuum
windows. In all cases, the effect is more important for giants than
for dwarfs.  Note that the dynamic range of both $H\gamma$ indices
is considerably lower than that of the two other Balmer lines. This is
because of the placement of the continuum passbands very close to the line
core, which artificially reduces the EW when $H\gamma$ is very strong,
and the inclusion of strong metal lines in the index passband, which
offset the decline of $H\gamma$ at low $T_{eff}$. $H\gamma_{HR}$ and
$H\gamma_{\sigma<130}$ behave similarly to one another as a function
of stellar parameters, even though they are defined and measured
in very different ways. 

\noindent {\it $<Fe>$ and Fe4383:} $<Fe>$ is the average of Fe5270 and
Fe5335. The behavior of the two latter indices as a function of stellar
parameters is essentially the same (Tripicco \& Bell 1995), so that
the discussion of $<Fe>$ applies to both Fe5270 and Fe5335.  All these
indices are very sensitive to {\it [Fe/H]}, in both giants and dwarfs:
their response to {\it [Fe/H]} is almost as strong as their response
to $T_{eff}$.  
Both indices are sensitive to gravity, being stronger for higher
gravities.

\noindent {\it $Mg\,b$ and Ca4227:} the $T_{eff}$-sensitivity of both
indices is very similar to the Fe indices. The most outstanding feature
of these indices is their strong response to $\log g$. They are almost
as sensitive to $\log g$ as they are to $T_{eff}$. The dependence on
$\log g$ is due to the effect of gas pressure on the damping wings:
the higher the surface gravity, the deeper the wings, and thus the index
is stronger. Ca4227 is mildly sensitive to {\it [Fe/H]} in both giants
and dwarfs. $Mg\,b$ is also mildly sensitive to {\it [Fe/H]} in dwarfs,
but displays a remarkable insensitivity to this parameter in the case
of giants. This result is in agreement with the results of Tripicco \&
Bell (1995), who showed that this index is predominantly sensitive to
$[Mg/H]$. Unfortunately, we lack this datum for the vast majority of the
library giants. This is a major shortcoming of all empirical spectral
libraries. In the case of the dwarfs, the mild sensitivity to {\it
[Fe/H]} can be explained by the fact that for $[Fe/H]>$--0.8 Fe and Mg
abundances vary in lockstep in disk stars (Edvardsson et al. 1993).

\section{Computing the Integrated Spectrum} \label{comp}

This section describes our procedures to compute model integrated
spectra. Two different procedures are followed: computation directly
from the the cluster's CMD (CMD-based) and those based on the theoretical
isochrones (isochrone-based). The latter are the basis for our discussion
in Paper II.

Figure \ref{fig4} shows the CMD of 47 Tuc out to a radius of $\sim$
30'', kindly provided by P. Guhathakurta (see Howell, Guhathakurta \&
Gilliland 2000).  This CMD was obtained with the Hubble Space Telescope
and is complete down to V=18.5 ($\sim$1 mag below turn-off). The CMD of
Figure \ref{fig4} is the starting point for our CMD-based computation
of the integrated spectrum of the cluster.

The computations can take at least two different routes: we can sum up
an {\it integrated spectrum} by selecting stars in the spectral library
to match each star in the CMD based on its atmospheric parameters,
or we can sum up particular {\it spectral indices}, based on fits to
the strengths of various spectral indiceas as a function of the stellar
parameters corresponding to each point of the CMD. The two methods should
yield exactly the same line index measurements. For this to occur,
the fitting functions must represent correctly the variation of the
spectral indices as a function of stellar parameters, and the spectral
library must sample adequately all the points in the ($T_{eff}$, $\log g$)
plane. Computation of the integrated spectrum is more straightforward and
also yields the entire spectrum, which allows a (very helpful) visual
comparison with the observations. However, it is not reliable if the
relevant stars are underrepresented. Integrated spectral indices are to
be preferred in that case, as fitting functions more easily extrapolate
and interpolate over missing stars. In the present case, we have noted
that the spectral library lacks giants that are metal-poor enough to
match 47 Tuc.  However we demonstrate in Section~\ref{res} that biases
in the computations due to sparseness of the spectral library are not
severe and can be readily corrected.

\subsection{CMD-based Integrated Spectrum}

As a first step, we computed a fiducial for the cluster by averaging
$(B-V)$'s in steps of 0.05 mag in V. Blue stragglers were averaged
in coarser steps of 0.2 mag. On the horizontal branch, we averaged
V magnitudes in steps of 0.02 mag in $(B-V)$. The number of stars
entering the average for a given fiducial point is a key quantity for the
computation of the integrated spectrum ($\mathcal{N}_i$ in Eq. 7 below).
Computing the integrated spectrum from the full CMD on a star-by-star
basis or from the fiducial yielded exactly the same spectrum and color.
Next, we derived dereddened absolute magnitudes and colors adopting
$E(B-V)$=0.04 and $(M-m)_0$=13.33 (Hesser et al. 1987). The fiducial
of 47 Tuc in the CMD was then transformed to the ($T_{eff}$, $\log g$)
plane, adopting the following set of calibrations: for the dwarfs,
we converted $(B-V)$ into $T_{eff}$ adopting the calibration by Alonso
et al. (1996) for {\it [Fe/H]}=--0.7. This calibration is a good match
to the $T_{eff}$s and $(B-V)$ of the library dwarfs, as shown in Figure
\ref{fig1c}. Surface gravities were determined from bolometric magnitudes,
masses and $T_{eff}$s, by taking the bolometric corrections given by
Alonso et al. (1995) and a mass $vs.$ $T_{eff}$ interpolated in the
isochrone by Salaris and collaborators (see description below) for 11 Gyr
and {\it [Fe/H]}=--0.7. For the giants, $T_{eff}$s were determined from
our own calibration given in Eq.(3) for {\it [Fe/H]}=--0.7, and $\log g$s
were determined in the same way as for the dwarfs, adopting the bolometric
corrections by Alonso et al. (1999) and a mass of 0.85 M$_\odot$. The
latter has been chosen to be consistent with the predictions of Salaris
models for the metallicity and age of 47 Tuc (see Paper II). The choice
of mass has a minor impact on the uncertainty of surface gravities:
a $\pm$ 0.1 M$_\odot$ variation in mass (which corresponds to a $\mp$ 4
Gyr change in age, for fixed metallicity, according to Salaris isochrones)
translates into roughly a $\pm$ 0.05 dex change in $\log g$.

The computation of the integrated spectrum proceeds as follows. At each
point of the fiducial, stars from the spectral library are identified that
match most closely the ($T_{eff}$, $\log g$) pair of the fiducial for
a metallicity as close as possible to {\it [Fe/H]}=--0.7.  This is done
by first searching for stars inside a given limit tolerance on $T_{eff}$
regardless of {\it [Fe/H]} or $\log g$. The stars so selected are then
submitted to a tolerance criterion in {\it [Fe/H]}, and the surviving
set is, in the end, selected for their surface gravities. In each of the
iterations, if no star is found in the library inside the tolerance of
each parameter, the latter is enlarged until at least one star is found
which matches the desired set of parameters. If the tolerance in $T_{eff}$
exceeds $\pm$ 100 K, the routine searches for other stars with parameters
selected so that the average $T_{eff}$ of the selected stars is closer to
the desired value. The initial tolerance limits in $T_{eff}$, $\log g$ and
{\it [Fe/H]} are 20 K, 0.1 dex and 0.05 dex respectively. The tolerance
limit in $T_{eff}$ never exceeds 250 K. In $\log g$, the tolerance limits
end up being more generous in regions of the CMD that are more sparsely
populated by the stellar library, such as K dwarfs, subgiants and M
giants. The same applies for the metallicities of K dwarfs and M giants.

The stellar spectrum for the $j^{th}$ point of the fiducial is given by

\begin{equation}
F_j(\lambda) = {1\over C}\sum_{i=1,n_j} {f_i(\lambda)\over d_i^2}
\end{equation}

where $n_j$ is the number of library stars selected per fiducial
point, $f_i(\lambda)$ is the spectrum of the $i^{th}$ star and $d_i$
is the distance between the $i^{th}$ star and the fiducial point in 
stellar parameter space, given by:

\begin{equation}
d_i^2 = \Delta \theta_{eff,i}^2 \,\, + \,\, \Delta \log g_i^2\,\,+ \,\,
\Delta {\it [Fe/H]}_i^2,
\end{equation}

The quantity $C$ is a normalization factor given by

\begin{equation}
C = \sum_{i=1,n_j} {1\over d_i^2}
\end{equation}

As a check on the final parameters for each fiducial point, we show in
Figure \ref{fig5} a comparison between the actual fiducial parameters and
the average parameters of the library stars adopted at each point. From
this figure, it can be seen that the final average $T_{eff}$s are in
close correspondence with the $T_{eff}$s required by the fiducials. In
$\log g$ and {\it [Fe/H]}, because of the limitations of the library
just discussed, more important differences, of the order of 0.5 dex
are found. However, this is not critical in most cases, first because
$T_{eff}$ is the leading parameter in shaping the stellar spectrum, and
second, because the discrepancies in $\log g$ and {\it [Fe/H]} occur in
areas of the CMD with very low contribution to the integrated light (see
discussion in Section \ref{contrib}). The most important discrepancy is in
fact of smaller size, namely, that most of the fiducial points of giants
with $T_{eff}$s between 4000 and 5200 K are represented by stars with
$<[Fe/H]>\sim$--0.6, rather than --0.7. This slight mismatch is due to
the paucity of metal-poor giants in the library, which forces our routine
to adopt higher-metallicity stars along the giant branch.  The impact of
this effect on our computations will be discussed in Section \ref{modobs}.

The next step in the procedure is to add up the spectra along the
fiducial so as to produce the final integrated model spectrum.
The latter is given by

\begin{equation}
\mathcal{F}(\lambda) = \sum_{j=1,N} \phi_j\,\,\, F_j(\lambda) 
\,\,\,10^{-0.4M_j}
\end{equation}
where $N$ is the number of fiducial points, $F_j(\lambda)$ is given by
equations (4) and (5), and $M_i$ is equal to $M_B$ of the fiducial point
in case of the blue spectra and an interpolation between $M_B$ and  $M_V$
for the red region. For fiducial points brighter than the completeness
limit of the CMD ($V_i<18.5$), $\phi_j$ is given by

\begin{equation}
\phi_j = \mathcal{N}_j\times{\phi_{com}\over N_{com}}
\end{equation}
where $\mathcal{N}_j$ is the number of stars in the $j^{th}$
fiducial point, $N_{com}$ is the total number of stars in the CMD
which are brighter than the completeness limit, and $\phi_{com}$ is
the corresponding value for a power-law IMF, integrated between masses
$m(V_i=18.5)$ and $m_{max}$, which is the initial mass of the most evolved
stars still visible in the cluster.  The latter two mass values are taken
from the isochrone of Salaris and collaborators (described below) for
an age of 11 Gyrs and {\it [Fe/H]}=--0.7. Below the completeness limit,
the fiducial is replaced by the isochrone of Girardi et al. (2000) for
10 Gyrs and {\it [Fe/H]}=--0.7, and $\phi_i$ is given by a power-law IMF,
with exponent $x$ (where $x = 1.35$ corresponds to the Salpeter IMF).

Mass segregation is extreme in the core of 47 Tuc. Howell et al.  (2001)
found a slope of $x = -5.0$ for the stellar mass function within the
cluster core radius. As our integrated spectrum samples only the inner 30
arcsec of the cluster, we adopt this severely dwarf-depleted mass function
below the completeness level of our CMD. Because the stellar library does
not contain enough cool dwarfs, the isochrone has to be truncated at $m =
0.6$ M$_\odot$. We adopted fitting functions to estimate the effects of
including the lower part of the main sequence (down to 0.1 M$_\odot$)
in the computations. As a result, we found that line indices change by
less than 1 \% when stars of lower masses are considered, so that the
lower main sequence can be safely neglected. The integrated $(B-V)$ 
remains likewise unchanged.


%

\section{Results} \label{res}

The comparison of the model predictions with the observations will
be presented as follows: we first compare the full CMD-based model
spectrum with the observations. In the absence of any important
reliance on theoretical isochrones, this test reveals any limitations
of the spectral library adopted in the synthesis. We then investigate
limitations of two kinds: the afore-mentioned paucity of library stars
on the metal-poor giant branch, and the absence of library stars with
the abnormal C,N abundance ratios that characterize half of the stars in
47 Tuc and other GCs.  These effects are investigated using line indices
and fitting functions. The latter requires a spectrum synthesis approach
(from model photospheres), whose details are presented in the Appendix.


The observed spectra of 47 Tuc used in this work come from two
different sources. An integrated spectrum was obtained in the interval
$\lambda\lambda$3440--4780 {\AA} with a resolution of $\delta\lambda\sim$
2.6 {\AA} (FWHM) at the 1.5 m telescope at CTIO (for further details,
see Leonardi (2001) in preparation). The second spectrum was kindly
provided by S. Covino and covers a much larger spectral region,
$\lambda\lambda$3690--7570 {\AA}, but with somewhat lower S/N and
resolution (3.4 {\AA}). It was obtained with the 1.52 m telescope at ESO
(see Covino, Galletti \& Pasinetti 1995). Both spectra were obtained
from drift scans inside the core radius of the cluster, which is $\sim$
22 arcsec (Trager, Djorgovski \& King 1993).  Thus the spectra sample
approximately the same stellar population as represented by the CMD
of Figure \ref{fig4}, which was obtained from a region within $\sim$
30 arcsec of the cluster center.

\subsection{CMD-based vs. Observed Spectra: Pseudocontinuum Slope}
\label{modobs}

Figure \ref{fig6} shows the comparison between the integrated spectra
of 47 Tuc and the CMD-based model spectra. The blue and red integrated
spectra are normalized at pseudocontinua points at 4365 {\AA} and
5358 {\AA} respectively. The overall match is not exact.  First, the
model spectrum is redder than the observed one. This difference is more
severe in the blue spectrum, amounting to about 9\% at 400 {\AA} from the
normalization point. In the red this difference amounts to only 3\% at the
same distance from the normalization point. The differences are an
order of magnitude larger than what can be accounted for by dereddening
errors. Furthermore, the absorption lines are in general slightly too
strong in the model spectra, the effect again being bigger in the blue.

We suggest that the discrepancy in the slope of the pseudocontinuum
is mostly due to flux calibration uncertainties in the observed 47 Tuc
spectra. In support of our suggestion, we compare in Table~\ref{tbl-2} our
predictions for the observed $(B-V)_0$ of 47 Tuc. The observed dereddened
integrated color within an aperture of 1 arcmin was taken from Chun \&
Freeman (1979). In this table, $(B-V)_{CMD}$ corresponds to the summation
of the colors of all the stars in the CMD, plus the extension with the
Salaris isochrone below the completeness limit, assuming a Salpeter IMF;
$(B-V)_{CMD}^{st}$ corresponds to the integrated color obtained by summing
the average color of the library stars entering the synthesis for each
fiducial point.  As expected, $(B-V)_{CMD}$ agrees perfectly with the
observations. We also find that $(B-V)_{CMD}^{st}$ is in close agreement
with the observations.  The slight mismatch ($\sim$ 0.1 dex) between
the metallicity of the cluster and the metallicity of the giant stars
employed in the synthesis (Figure \ref{fig5}) accounts for less than 15\%
of the pseudocontinuum slope differences reported above. The excellent
agreement between the observed $(B-V)_0$ and $(B-V)_{CMD}^{st}$ shows
that the stars employed in the synthesis have the right overall color.

Since we are able to reproduce to high accuracy the integrated color
of the cluster using the colors of the library stars, and given that
the flux calibration of the spectral library is said to be free from
any systematic errors (see Vazdekis 1999 for a discussion), we conclude
that the mismatch between the spectrophotometric observations of 47 Tuc
and our model pseudocontinuum slope must be due to some error in the flux
calibrations of the observed 47 Tuc spectra. If so, the error is not
serious and we consider it no further.

\subsection{CMD-based vs. Observed Spectra: Absorption Lines} \label{lind}

In this section, we discuss the line indices presented in
Section~\ref{indstar}. As discussed in that section, the $H\gamma_{HR}$
and $H\gamma_{\sigma<130}$ indices behave in virtually the same way
as a function of stellar parameters, so that the results described
in this section apply to both indices.  However, as discussed in
Section~\ref{indstar}, the correction to the effect of anomalous
C,N abundances on $H\gamma_{HR}$, is very uncertain (see also
Section~\ref{synth} and Appendix), so that we will focus our discussion
on $H\gamma_{\sigma<130}$. Moreover, the latter is very relevant to our
purposes because the study of Vazdekis et al. (2001) on the spectroscopic
age of 47 Tuc relies heavily on this index.

Besides the overall disagreement in pseudocontinuum slope, line indices
are systematically overestimated in the model spectra.  This can be seen
by careful assessment of individual features in Figure \ref{fig6}, but
can be better appraised in Figures \ref{fig6a} and \ref{fig6b}, where
the models and observations are compared in a number of Balmer line
$vs.$ metal-line plots. In all these plots, the filled pentagons with
the error bars represent the observed 47 Tuc indices, while the open
circles represent the indices in the CMD-based model spectra. Indices
were measured in model spectra, rebinned and convolved to the dispersion
and resolution of the two sets of observed spectra. The measurements in
both observed and model spectra are listed in Table~\ref{tbl-4}. For the
red observed spectrum, errorbars are taken from Covino et al. (1995),
and for the blue, they are estimated from the observations by considering
only the error in the wavelength scale, given the extremely high S/N of
this spectrum.

The results from these comparisons can be summarized as follows:

1) $H\delta$ and $H\beta$ are well matched by the CMD-based model
whereas the model overpredicts $H\gamma$ by more than 1 $\sigma$;

2) Among the metal-line indices, $<Fe>$ is well matched by the model,
$Mg\,b$ is roughly matched, but Fe4383 and Ca4227 are overpredicted,
especially the latter. The differences are greater in the blue indices,
and this trend is confirmed by visual inspection of other lines (Figure
\ref{fig6}).

Only three basic ingredients enter the CMD-based synthesis: an assumed
metallicity for 47 Tuc, the spectral library and its stellar parameters,
and the adoption of an isochrone with a Salpeter IMF to account for lower
main sequence stars below the completeness limit of the input CMD. Since
the faint lower main sequence stars have a negligible impact on the line
index predictions, even for wide variations of the power-law IMF exponent,
the first two ingredients are the key ones, and we explore them further.

In Section \ref{comp} we pointed out that the average metallicity
of giants employed in the synthesis of the integrated spectrum is
higher by $\sim$ 0.1 dex than the estimated metallicity of 47 Tuc ({\it
[Fe/H]}=--0.7). This is due to the paucity of metal-poor giants in the
spectral library adopted. In order to estimate the impact of this error,
we calculate the effect of decreasing the average metallicity of the
input giants by 0.1 dex, by using line-index fitting functions.  This is
a good way of estimating changes in model predictions due to fine-tuning
of input parameters because fitting functions smooth over the effects of
sparseness of the spectral library in certain regions of stellar parameter
space. The fitting functions adopted were those of Worthey et al. (1994)
for Fe4383, Ca4227, $Mg\,b$, and $<Fe>$, Schiavon (2002, in preparation)
for $H\beta$ and $H\delta_F$, and Concannon (2001, private communication)
for $H\gamma_{\sigma<130}$.  We note that the fitting functions of Worthey
and collaborators are built for an EW system that is different from
ours, which is based on much higher-resolution spectra. For this reason,
the use of those fitting functions will be kept strictly differential,
in a percentage sense

Our procedure was to vary the input metallicity of giant stars by
--0.1 dex and then infer the percentage change in each line index. The
latter was then applied to the model predictions. The results are
displayed as the dotted arrows in Figures \ref{fig6a} and \ref{fig6b}.
The direction of the arrows can be easily verified from the discussion in
Section~\ref{indstar}. Basically, all corrections are small.  $H\beta$,
$H\gamma$ and all metal-line indices become stronger for more metal-rich
giants. Thus, correcting for the effect of too metal-rich giants in the
library decreases these indices. The opposite happens to $H\delta_F$
because it is the only index that becomes weaker in the spectra of more
metal-rich giants.  After correcting for these effects, the agreement
of model predictions with observations is slightly worse in the case
of $H\beta$ and $H\delta_F$, although formal agreement for $H\beta$
is preserved, and for $H\gamma_{\sigma<130}$ the agreement is actually
improved. The changes to metal lines are generally in the right direction
but not large enough to remove all discrepancies.

The next possible source of error is the metallicity of 47 Tuc itself.
The value adopted so far has been taken from Carretta \& Gratton (1997),
and it relies on a detailed abundance analysis of high resolution spectra
of cluster stars. The uncertainty in {\it [Fe/H]} quoted by those
authors is $\sim$ 0.1 dex. Thus, it is worth testing the sensitivity
of our predictions to a slight decrease in the assumed {\it [Fe/H]} of
47 Tuc. The procedure in this case is a bit trickier, since the effect
of the metallicity on the $T_{eff}$s (which are inferred from stellar
colors) also needs to be taken into account. For a given $(B-V)$, the
inferred $T_{eff}$ is slightly lower, because of the blueing effect
of a lower metallicity. We infer new sets of stellar parameters from
colors and absolute magnitudes by using the same calibrations as those
described in Section \ref{comp}, for the slightly lower metallicity {\it
[Fe/H]}=--0.75. The new values are then input into the fitting functions
to obtain new percentage changes in the line indices.

The results are displayed in Figures \ref{fig6a} and  \ref{fig6b} in the
form of solid arrows. The indices corrected for both the metallicity of
the giants and 47 Tuc are listed in Table~\ref{tbl-4}. For Balmer line
indices, the response is stronger than in the previous test, for the solid
arrows have about the same length as the dotted arrows, even though the
former were computed for a change in metallicity which is only half as
large. This is because the new $T_{eff}$s are lower than those obtained
when assuming {\it [Fe/H]}=--0.7, for the reasons explained above. For
metal-line indices, the change depends on the balance between $T_{eff}$
and {\it [Fe/H]}-dependence.  From these figures it can be seen that, when
we combine the correction for the higher metallicity of the library giants
with a putative 0.05 dex decrease in the assumed metallicity of 47 Tuc,
we achieve an overall better compromise with the observations. Three of
the four metal lines and two out of three Balmer lines are now in good
agreement with the observations. Ca4227 is still wildly overpredicted,
however, and $H\delta_F$, which formerly agreed, is now too strong,
because of its dependence on metallicity, as explained above. 

In summary, overall agreement between the CMD-based model and the
observations is improved by correcting for the lack of metal-poor
giants in the library and by assuming that 47 Tuc is a little more
metal-poor than the standard value ({\it [Fe/H]}=--0.75, yet safely
within the errorbar of Carreta \& Gratton 1997).  In the next section,
we investigate a final cause for the remaining mismatches in $H\delta_F$
and Ca4227, namely, the effect of the abnormal strength of CN and CH
lines in the integrated spectrum of the cluster.

\subsection{Spectrum Synthesis from Model Atmospheres: Assessing the
Effect of CN-strong Stars on Integrated Line Indices} \label{synth}

The $\lambda\lambda$3800--4400 {\AA} interval is filled with lines
due to CN and CH. In particular, both the passbands and continuum windows
of $H\delta_F$, $H\gamma$ and Ca4227 are heavily blended by many lines
due to CN and CH (Castilho 1999).  It is well known that about half
the stars of 47 Tuc and other GCs present an unusual abundance pattern,
characterized by an abnormally high nitrogen abundance and a mild carbon
underabundance (Cannon et al. 1998 and references therein). The net effect
is to enhance the opacity due to the CN molecule and decrease that due
to CH, as compared to stars with solar abundance ratios. Our spectral
library lacks stars with abundance ratios similar to the ones found in
the CN-strong stars in 47 Tuc. Therefore, we resort to spectrum synthesis,
based on state-of-the-art model atmospheres and a comprehensive database
of atomic and molecular data, in order to investigate the origin of the
discrepancies discussed in the previous sections.

Computing reliable synthetic spectra from model atmospheres of cool stars
is an extremely difficult task, involving the accurate determination of
hundreds of thousands of oscillator strengths, even for a relatively
narrow spectral region like the one under study in this paper. In the
blue, this task is even harder, given the crowding of atomic lines even
in the highest resolution attainable in stellar spectroscopy. However,
although we may not be able to produce a perfect absolute match to
the observations, it should be possible to use the model spectra
differentially, to investigate the effect of changes in abundance ratios.

We computed synthetic spectra for the atmospheric parameters of a typical
giant, subgiant and turn-off star in 47 Tuc. The stellar parameters
adopted were: ($T_{eff}$, $\log g$) = (4750 K, 2.50) for a red giant,
(6000 K, 4.0) for a turn-off star and (5250 K, 3.0) for a subgiant.
In the absence of model atmospheres computed for the exact metallicity of
47 Tuc, all spectra were computed for {\it [Fe/H]}=--0.5. Therefore,
this analysis must be kept strictly differential. Here we concentrate on
the main results leaving the details for the Appendix.  The synthetic
spectra were calculated using two different abundance patterns: one
for solar chemical composition (Grevesse \& Sauval, 1998), scaled to
the adopted $[Fe/H]$=--0.5 dex, and the other for the abundance pattern
of CN-strong stars in 47 Tuc, namely, $[C/Fe]$=--0.2 and $[N/Fe]$=+0.8
(Cannon et al. 1998).

The spectral differences resulting from these abundance patterns
are illustrated in Figure \ref{fig13}.  In Figures \ref{fig14} and
\ref{fig14b} we display a zoom of Figure \ref{fig13} in the regions of
$H\gamma$ and $H\delta$ respectively. While the CN bands at $\sim$
4160 {\AA} are stronger in the synthetic spectrum computed with
a non-solar abundance pattern, the G band, at $\sim$ 4300 {\AA}
(due mainly to CH) is weaker.  The differences are large for giants and
subgiants, while for turn-off stars they are far less important. This
is because CN and CH lines are much weaker at the high $T_{eff}$ and
$\log g$ characteristic of turn-off stars.

We checked whether the predicted model differences in CH and CN band
intensities match those measured in spectra of stars from 47 Tuc. For
that purpose, we measured the indices C(4142) and $s_{CH}$ as defined
respectively by Norris \& Freeman (1979) and Briley et al. (1994) in the
synthetic spectra of the CN-normal and CN-strong giant model. According
to our computations, the CN-strong giant is stronger in C(4142) by 0.15
mag and weaker in $s_{CH}$ by 0.05 mag. These variations are consistent
with the differences measured among giant stars in 47 Tuc by Norris \&
Freeman and Briley et al.

A look at Figures \ref{fig13}, \ref{fig14} and \ref{fig14b} allows
a quick estimate of how the line indices should change as a function
of relative CN/CH line strength. In the case of Ca4227, the index is
much weaker in CN-strong stars because of the severe contamination of
the blue continuum window by CN lines. The effect on Balmer lines is
more subtle. $H\gamma_{\sigma<130}$ is slightly weaker in the spectra
of CN-strong stars because the central passband, which includes the
metallic feature at $\sim$ 4350 {\AA}, which is due to a blend of
FeI, FeII, TiII, CrI and MgI lines, but is also strongly contaminated by
CH lines, which are weaker in those stars. In the case of H$\delta_F$,
CN lines contaminate both continuum and central passbands, so that
the net effect on the measured index is not large. For all indices,
the effect in turn-off stars is far less important, given the relative
weakness of molecular lines in their spectra.

To assess the net effect of CN-strong stars on the CMD-based model, we
summed the spectra displayed in Figure \ref{fig13} adopting the weights
given in Table~\ref{tbl-3}.  In this way, we produced two integrated
spectra, one comprised of only CN-normal stars, and the other comprised
of a mixture of half CN-normal and half CN-strong stars.  Obviously, the
absolute values of such integrated spectra do not reproduce accurately
the observations because by construction they do not sample the full
range of atmospheric parameters characteristic of cluster stars, and
also because they were computed with a metallicity slightly higher
than 47 Tuc's. However, they do provide a rough measure of the effect
of CN-strong stars on the modeled indices. The synthetic spectra were
then smoothed and rebinned to the same resolution and dispersion as the
blue observed spectrum, and indices were measured in the same way as in
the observed and previous model spectra.

The resulting corrections are shown as the arrows labeled ``CN'' in
Figures \ref{fig6a} and \ref{fig6b}. The corrected indices are listed
in Table~\ref{tbl-4}. To compute these, we have measured the percentage
change in each index caused by CN-strong stars in the model and applied
the same percentage shift as a further correction to the previous
best model presented in the last section.  We find that $H\delta_F$
is increased by $\sim$ 10\% when CN-strong stars are included, so
that the corrected $H\delta_F$ is stronger than the measured values
by $\sim$ 0.4 {\AA}. However, we do not put too much weight on this
discrepancy, given that our synthetic spectrum seems to underpredict
this line as well as a few moderately strong metal lines included
in the passband of the index, at $\sim$ 4110--4112 {\AA} (see Figure
\ref{fig11} in the Appendix). Because the total absorption inside the
index passband is underestimated, the percentage change due to the
variation in CN line-strength may be overestimated. Most importantly,
the final correction to the line index results from a delicate balance
between the corrections due to extra CN absorption both in the continuum
windows and index passband. In a spectral region heavily contaminated
by CN lines, both corrections (to continuum and passband) are large
(see Figure \ref{fig14b}), and the resulting correction to the line
index is small, so that small errors in the line strengths of CN lines
in either the continuum and passband would result in a sizably wrong
final correction to the index. In particular, if the CN-line strengths
in the index passband were slightly underestimated relative to CN-line
strengths in the continuum windows, the present index correction would
go in the wrong sense.


%

Unfortunately the wavelength interval of the synthetic spectrum does
not include the Fe4383 red continuum window. However, it can be seen in
Figure \ref{fig14} that CH line variation is very mild in the spectral
vicinity of the index, so that it is unlikely that Fe4383 is strongly
affected by these lines, and we chose not to make any correction to
model predictions of Fe4383. In the case of $H\gamma_{\sigma<130}$ we
find that the index is reduced by $\sim$5\% when CN-strong stars are
included in the synthesis, because of the aforementioned contamination
of the index by CH lines. This is an artifact of the index definition,
as illustrated in Figure \ref{fig14}, where we display a zoom of the
spectra from Figure \ref{fig13} in the region of $H\gamma$. The continua
and passband of $H\gamma_{\sigma<130}$ are shown in the upper panel. From
the figure, it is clear that the reduction in $H\gamma_{\sigma<130}$
when considering CN-strong (CH-weak) stars is due to the reduction in
the intensity of the CH feature at $\sim$4352 {\AA}, which is included
in the index passband. This result should be taken as a warning, for it
means that $H\gamma_{\sigma<130}$ is affected, though mildly, by carbon
and nitrogen abundance ratios, which may differ between giant ellipticals
and the Milky Way (Trager et al. 1998).

It is interesting to note that the results of Gibson et al. (1999),
Jones (1996) and Rose (1994) for CaI were opposite to ours, in the
sense that the index was {\it under}predicted by their models. This is
an artifact of the differing index definitions adopted in the different
works. In particular, the other authors adopted high-resolution, very
narrow-band indices, whose continuum windows were much less affected
by CN/CH opacities than Ca4227, whose blue continuum contains strong CN
lines. The latter make the index much weaker in the observed spectrum,
due to the lowering of the blue continuum by the stronger CN lines in
the spectra of CN-strong stars from 47 Tuc (see next section). All
in all, we stress the fact that our correction for the intensity of 
CN lines brings our predicted Ca4227 essentially into agreement with
the observations.

A similar study was performed by Tripicco \& Bell (1992), where
integrated synthetic spectra of purely CN-strong and CN-weak stars have
been computed and compared to the integrated spectrum of 47 Tuc. The
authors found that a better fit was obtained when a purely CN-strong
integrated spectrum is employed. This is in disagreement with our results
shown in Figures \ref{fig6a} and \ref{fig6b}, because if we assume that
the center of 47 Tuc is made of 100\% of CN-strong stars, the corrections
for the effect of these stars in the Figures would drive our model
predictions far from the observations. Tripicco \& Bell (1992) claim that
their result is consistent with the evidence, found by Norris \& Freeman
(1979), of an increase of the ratio of CN-strong/CN-weak fraction towards
the cluster center. Indeed, Briley (1997) found further evidence, on the
basis of better statistics, suggesting that the fraction of CN-strong
stars in the inner 10 arcminutes of the cluster is $\sim$ 65\%. However,
the data in Briley (1997) do not seem to suggest that the proportion
of CN-strong stars increases all the way into the cluster center, but
rather keeps a constant value from 4 to 10 arcminutes from the center,
undergoing a sharp decline towards larger radii. It is nevertheless very
hard to assess the reasons for the disagreement between Tripicco \& Bell
and our work, given the many differences between the two procedures,
such as: the abundance patterns adopted in the computations of stellar
synthetic spectra are slightly different; their integrated synthetic
spectrum is based on a theoretical isochrone, while ours is based on
an observed color-magnitude diagram; they compute synthetic spectra
in a fine grid of stellar parameters, while we just infer a somewhat
cruder estimate of the corrections to the model integrated spectrum from
computations for a much coarser grid of stellar parameters. A resolution
of this impass awaits the production of a finer grid of synthetic spectra
with updated model atmospheres and opacities (Castilho \& Barbuy 2002,
in preparation).

In summary, we conclude that considering the effect of CN-strong stars in
the integrated spectrum of 47 Tuc brings our model predictions into better
agreement with the observations, in particular for Ca4227, for which
the corrected model prediction is almost exactly correct.  The case
of $H\delta_F$ is more complicated, because, as opposed to $H\gamma$,
our synthetic spectrum is not a good match to the spectra of normal cool
giants in the region of this index, as shown in the Appendix. Comparing
the results from the three Balmer lines under study, we achieve formal
agreement with the observations for both $H\gamma_{\sigma<130}$ and
$H\beta$, while $H\delta_F$ is overpredicted by $\sim$ 0.4 {\AA}. The
residual mismatches in $H\gamma_{\sigma<130}$ and $H\beta$ are 0.03
and 0.09 {\AA} respectively. Such mismatches translate into age errors,
according to the models of Paper II, of +4 Gyrs for $H\delta_F$, --1.5
Gyr for $H\beta$ and --1 Gyr for $H\gamma_{\sigma<130}$.

\subsection{Uncertainties due to Input Stellar Parameters} \label{unc}

As noted in Section \ref{dwarfs}, the $T_{eff}$ and metallicity scales
of metal-poor stars from different sources in the literature may differ
by up to 100 K and 0.1 dex. Scale uncertainties of this magnitude are
present in the adopted stellar parameters of our stellar library. We ask
here how such uncertainties would affect our CMD-based model spectrum,
if at all. Consider $T_{eff}$ errors first. It turns out that these
completely cancel out. This is because the calibration adopted to
convert $(B-V)$s from the CMD of 47 Tuc to $T_{eff}$s was inferred from
the $T_{eff}$s and colors of the library, in the case of the giants,
while for the dwarfs the calibration adopted is fully consistent with
the $T_{eff}$s and $(B-V)$s of the library stars. Thus, it does not
matter which parameter is formally used to select stars from the spectral
library: both $T_{eff}$ and $(B-V)$ produce the same result for a fixed
{\it [Fe/H]}. $T_{eff}$ is just an intermediate placeholder, and the
real calibration taken from the library is line-strength vs. $(B-V)$,
which are both well measured empirically. Therefore, the effect of any
error in the $T_{eff}$-scale of the library stars is negligible.

The same is not true of {\it [Fe/H]} errors. Any systematic error in the
{\it [Fe/H]}-scale of the library stars will produce a mismatch between
observed and model line indices. The error in the metallicity which
is deduced from the integrated line indices is precisely equal to the
scale zeropoint error in the {\it [Fe/H]}-scale of the stellar library.
In fact, we seem to have detected such a mismatch, in that 47 Tuc
appears to be a little more metal-poor (by $\sim$ 0.05 dex) according
to our synthesis than the previous high-resolution analysis. In fact,
we cannot tell from our present knowledge whether the mismatch is an
error in the high-resolution analysis (which we have assumed until now)
or in our own stellar library {\it [Fe/H]} zeropoint. In either case,
the effect is reassuringly small.

Finally, we note that the temperature cancellation effect that is present
in the CMD-based model is {\it not} achieved when making models based
on theoretical stellar isochrones, where stars in the synthesis are
ultimately selected by $T_{eff}$, not $(B-V)$. This problem will be the
subject of a detailed discussion in Paper II.

Another source of error in the computation of the CMD-based model spectrum
is the assumed reddening. If the reddening is overestimated, dereddened
colors are bluer, so that the stars selected in the synthesis tend to be
hotter (for a fixed {\it [Fe/H]}), leading to stronger Balmer lines and
weaker metal lines in the CMD-based model spectrum.  The value adopted
in this work, $E(B-V)=0.04\pm0.01$, was taken from Hesser et al. (1987).
We estimated the uncertainties in the model predictions due to errors in
reddening by varying the input $E(B-V)$ by $\pm$ 0.01 mag and measuring
all line indices in the model spectra obtained. The results are shown as
the arrows labeled ``E(B-V)'' in Figures \ref{fig6a} and \ref{fig6b}.
From the size of these arrows, and with the aid of the isochrone-based
models presented in Paper II, we estimate that errors due to reddening
cause an uncertainty of 1 Gyr in the spectroscopic age inferred from
Balmer lines. This result is fully consistent with the uncertainties
induced by reddening in the ages inferred from the fit to the turn-off
absolute magnitude.

\subsection{Fractional Contribution to the Integrated Light from Different
Evolutionary Stages} \label {contrib}

\begin{figure}[htpb]
\plotone{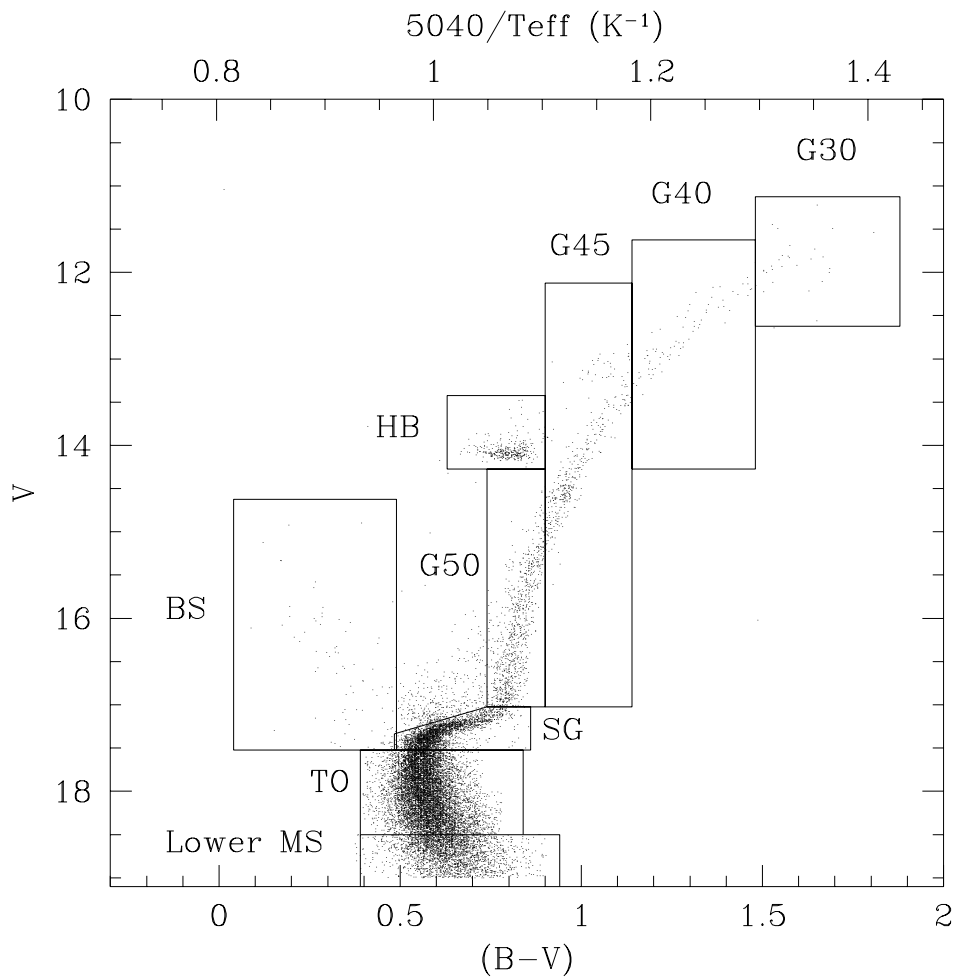} \caption{CMD of 47 Tuc out to
a radius of $\sim$ 30'', from Howell et al. (2000).  The boxes define
different evolutionary stages, whose fractional contribution to the
integrated light is displayed in Table~\protect\ref{tbl-3}.  Notice that
the subgiant branch (SG) has been defined so as to eliminate stars that
are too bright, thus minimizing effects due to crowding and binarity.
When taken into account, those stars contribute negligibly both to
the integrated light and line indices.  The scale on top provides
$\theta_{eff}$ for giant stars, given by equation (3).
\protect\label{fig4} } 
\end{figure}

\begin{figure}[htpb]
\plotone{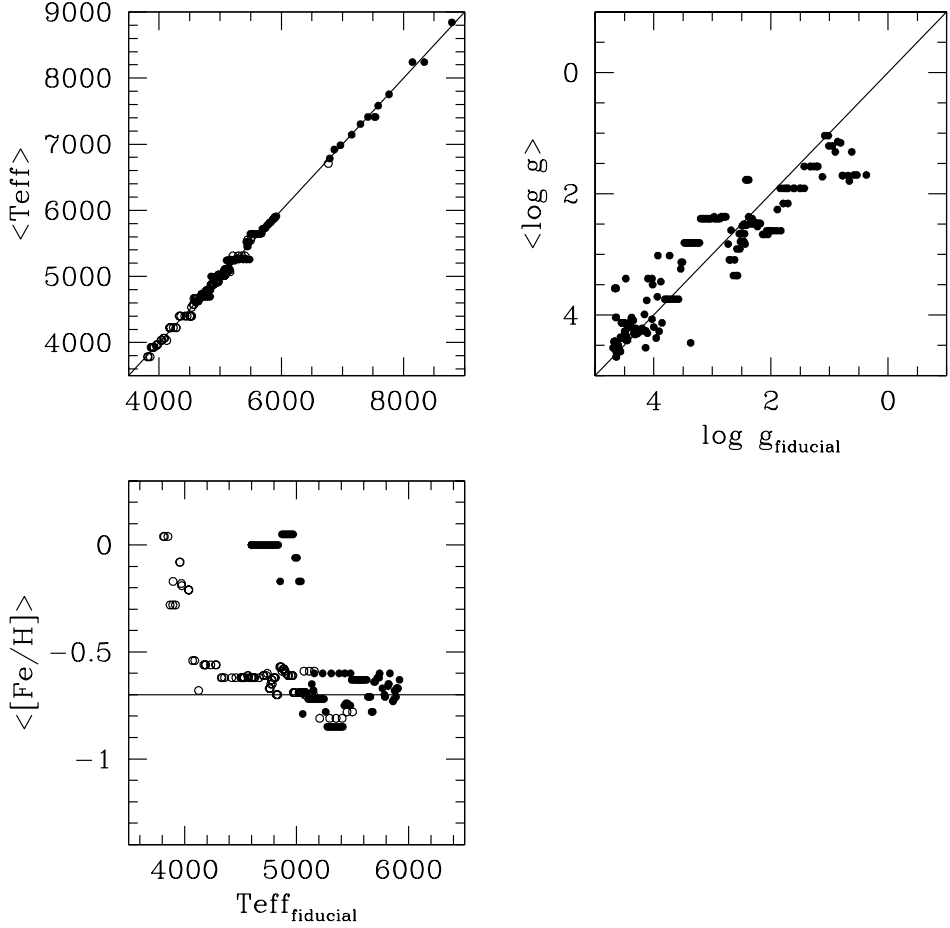}
\caption{Comparison between the stellar parameters of the 47 Tuc fiducial
and the average parameters of the stars actually used as input in the
CMD-based synthesis.  Open circles represent stars with $\log g <$3.6,
and closed circles represent stars with higher gravities. The nominal
$[Fe/H]$ of 47 Tuc, equal to --0.7, is indicated as the horizontal
line in the lower-left panel.  $T_{eff}$s are very well reproduced by
the library. Differences in $\log g$ are negligible.  The very deviant
points in {\it [Fe/H]} are M giants and K dwarfs. K dwarfs contribute
little to the integrated light, while for M giants metallicities are 
extremely uncertain. The $\sim$ 0.1 dex difference in {\it [Fe/H]} for
giants with $T_{eff}$s between 4000 and 5200 K is in fact more important
(see text).
\protect\label{fig5} } 
\end{figure}

\begin{figure}[htpb]
\plotone{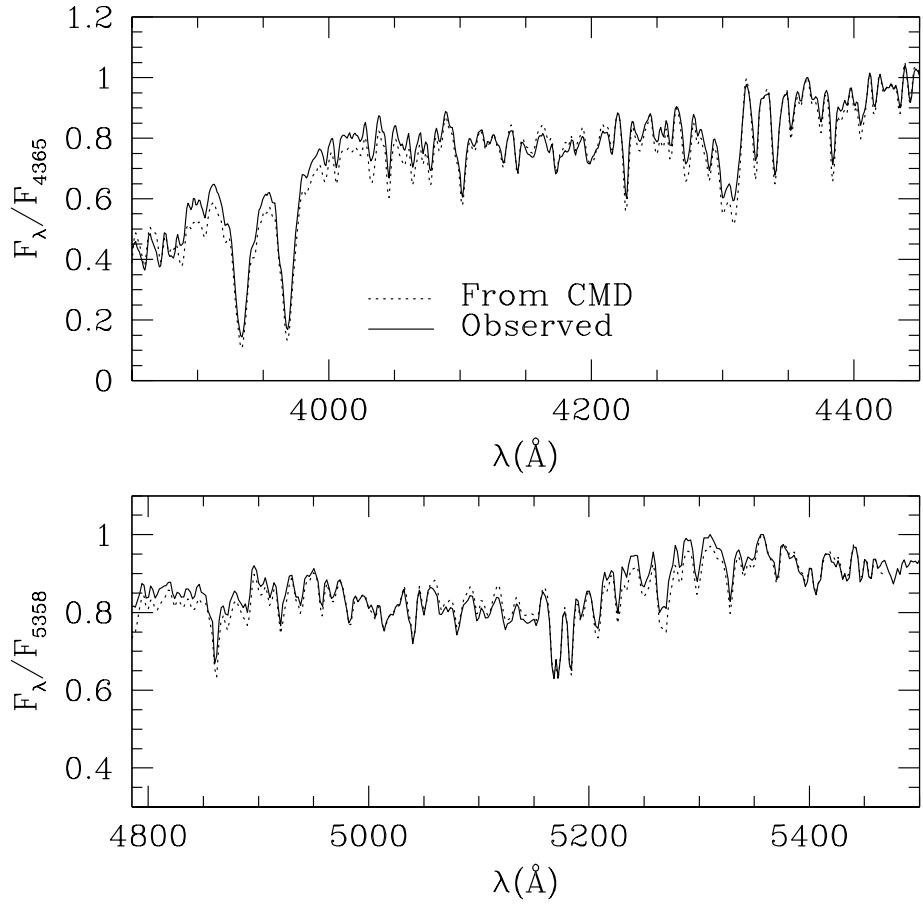}
\caption{Comparison between the model spectra for 47 Tuc computed from
the CMD and observed spectra by Leonardi (2001) (upper panel) and Covino
et al. (1995) (lower panel).  
\protect\label{fig6} } 
\end{figure}

\begin{figure}[htpb]
\plotone{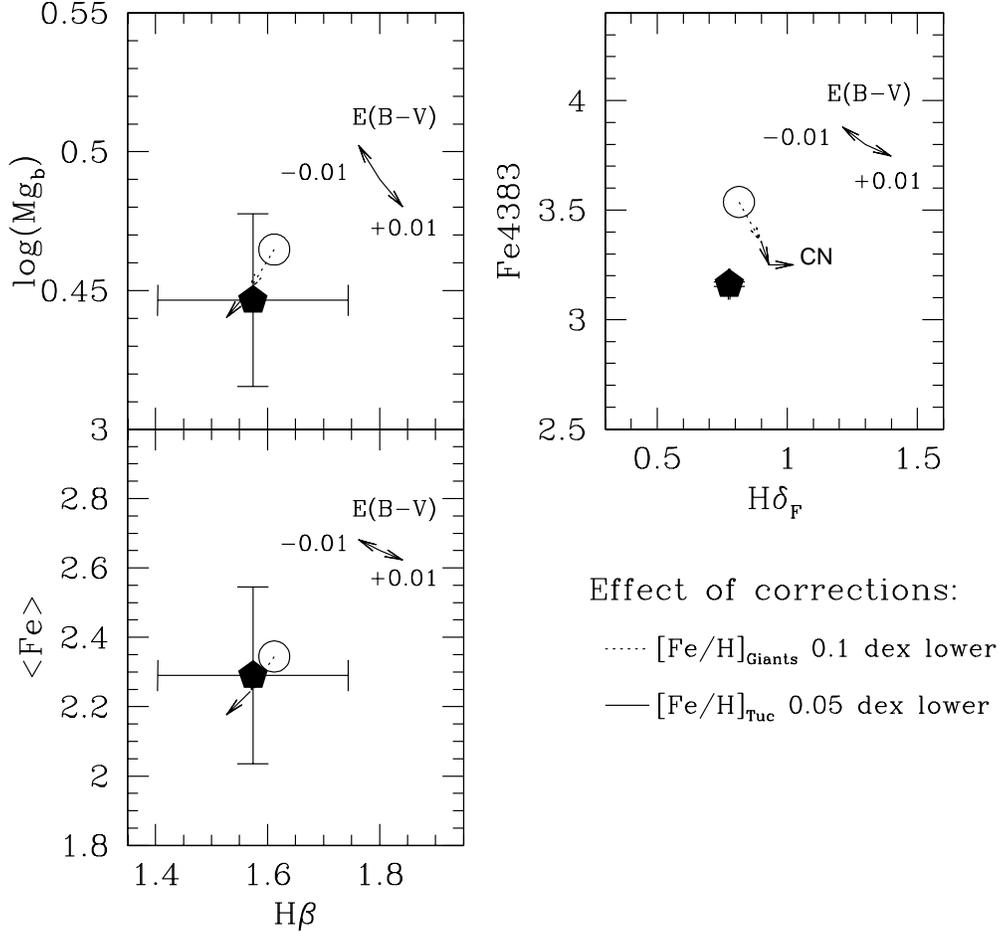}
\caption{Comparison between model predictions and observations for various
Balmer and metal-line indices. Observations are represented by the filled
pentagons with errorbars.  The open circles represent the CMD-based model,
including blue stragglers. Arrows labeled with ``E(B--V)'' indicate how
model predictions change if the assumed reddening changes by $\pm$ 0.01.
The arrow labeled ``CN'' in the right panel indicates how $H\delta_F$
changes after correcting for the presence of CN-strong stars.  No such
correction was applied to Fe4383 because the density of CH lines in the
vicinity of this index appears to be negligible. {\it Dotted arrows}
show how the model changes when the metallicity of the giants adopted
in the synthesis is decreased by 0.1 dex. {\it Solid arrows} show the
additional correction that results from assuming that the metallicity
of 47 Tuc is 0.05 dex lower than the standard value, {\it i.e.} {\it
[Fe/H]}=--0.75. Applying both corrections brings the metal lines into
better overall agreement without disturbing $H\beta$ unduly, but at the
risk of overpredicting $H\delta_F$.
\protect\label{fig6a}
} 
\end{figure}

\begin{figure}[htpb]
\plotone{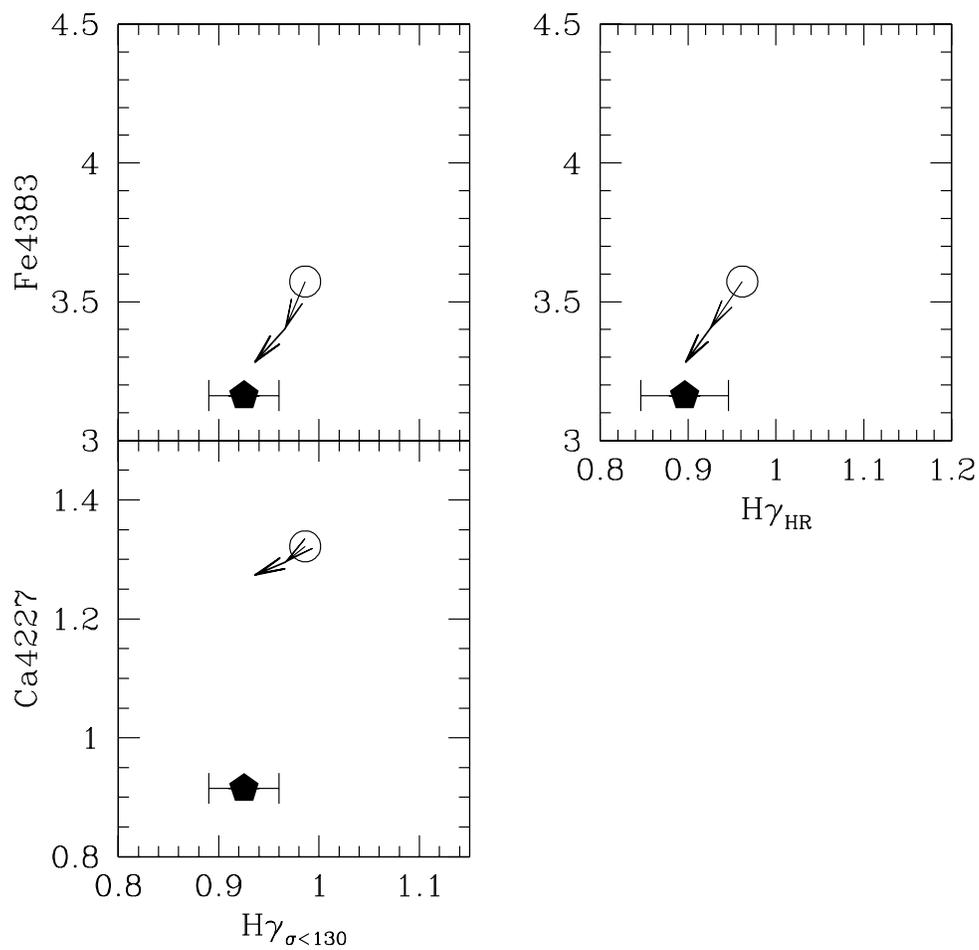}
\caption{Same as Figure \protect\ref{fig6a} but for $H\gamma_{\sigma<130}$
and two metal indices. Applying both corrections described in Figure
\protect\ref{fig6a} and in the text improves $H\gamma_{\sigma<130}$
and Fe4383, but neither correction is able to reconcile Ca4227 with the
observations.  ``CN'' indicates the estimated corrections for the presence
of CN-strong stars. Accounting for the latter effect essentially removes
the whole discrepancy in Ca4227 without perturbing $H\gamma_{\sigma<130}$
unduly.  All other arrows are labeled according to Figure \ref{fig6a}.
\protect\label{fig6b} } 
\end{figure}

\begin{figure}[htpb]
 \plotone{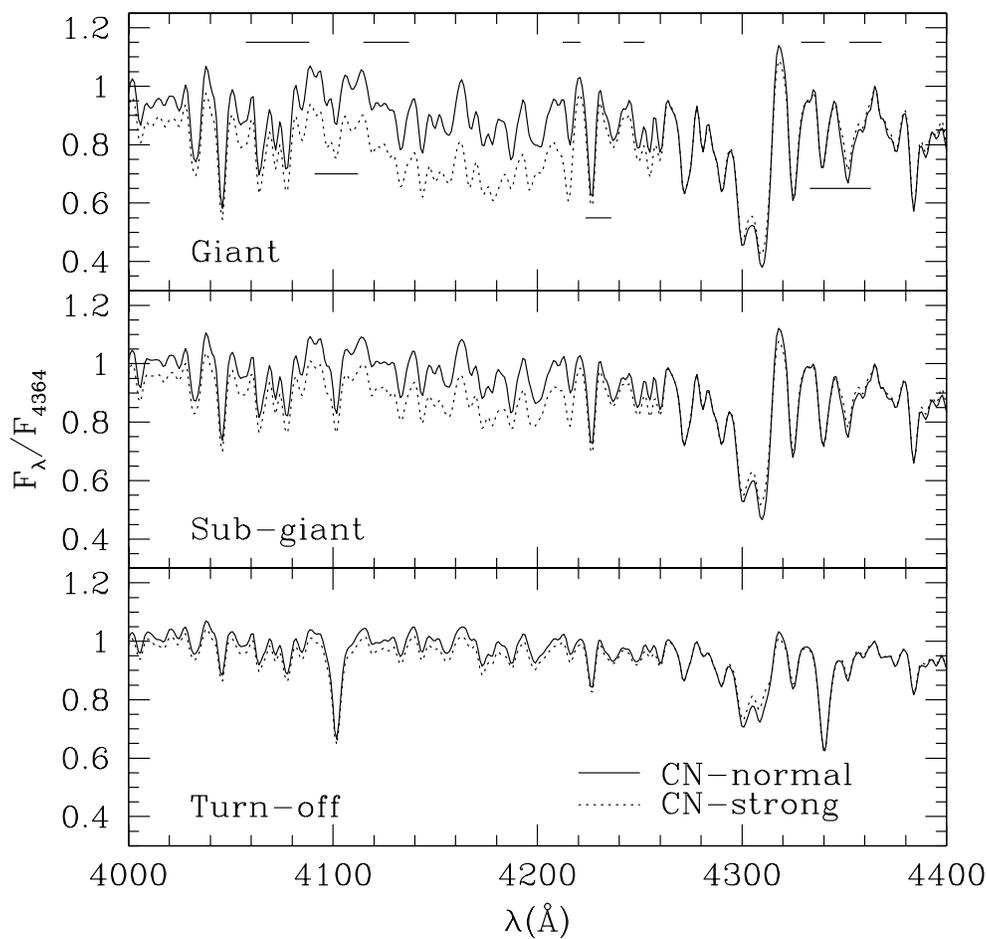} \caption{Synthetic
spectra showing the effect of C,N abundance variations for stars of
representative evolutionary stages in 47 Tuc. Notice that CN-strong
stars have stronger CN lines around $\sim$4160 {\AA} but a
weaker G-band at $\sim$ 4300 {\AA}. The passbands of $H\delta_F$,
Ca4227 and $H\gamma_{\sigma<130}$ are shown in the top panel. Notice
the contamination of the blue continuum window of Ca4227 by a strong CN
line, which makes the index weaker in the spectra of CN-strong stars.
\protect\label{fig13} } 
\end{figure}

\begin{figure}[htpb]
\plotone{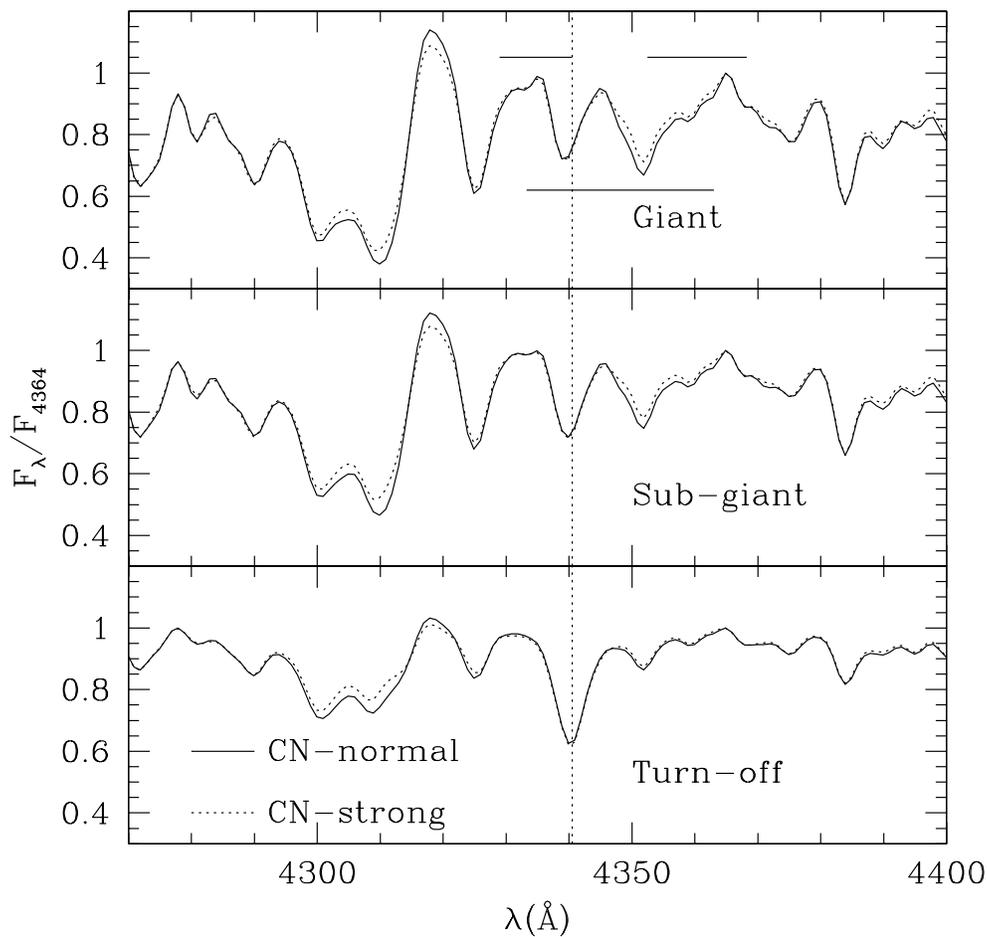}
\caption{Same as Figure \protect\ref{fig13}, in the region of $H\gamma$.
The upper panel shows the passbands of the $H\gamma_{\sigma<130}$ index.
The vertical line indicates the central wavelength of $H\gamma$. Notice
the presence of CH lines at $\sim$ 4350 {\AA}, which are weaker in
the spectra of CN-strong stars. As they are included in the central 
bandpass of $H\gamma_{\sigma<130}$, the index is stronger in CN-normal
stars. Note further that even the main H$\gamma$ feature at 4340 {\AA}
contains a metal line which causes the center of the blended feature to
walk to the blue of 4340 {\AA} in the spectrum of the cool giant.
All in all, $H\gamma$ is a complicated feature.
\protect\label{fig14}
}
\end{figure}

\begin{figure}[htpb]
\plotone{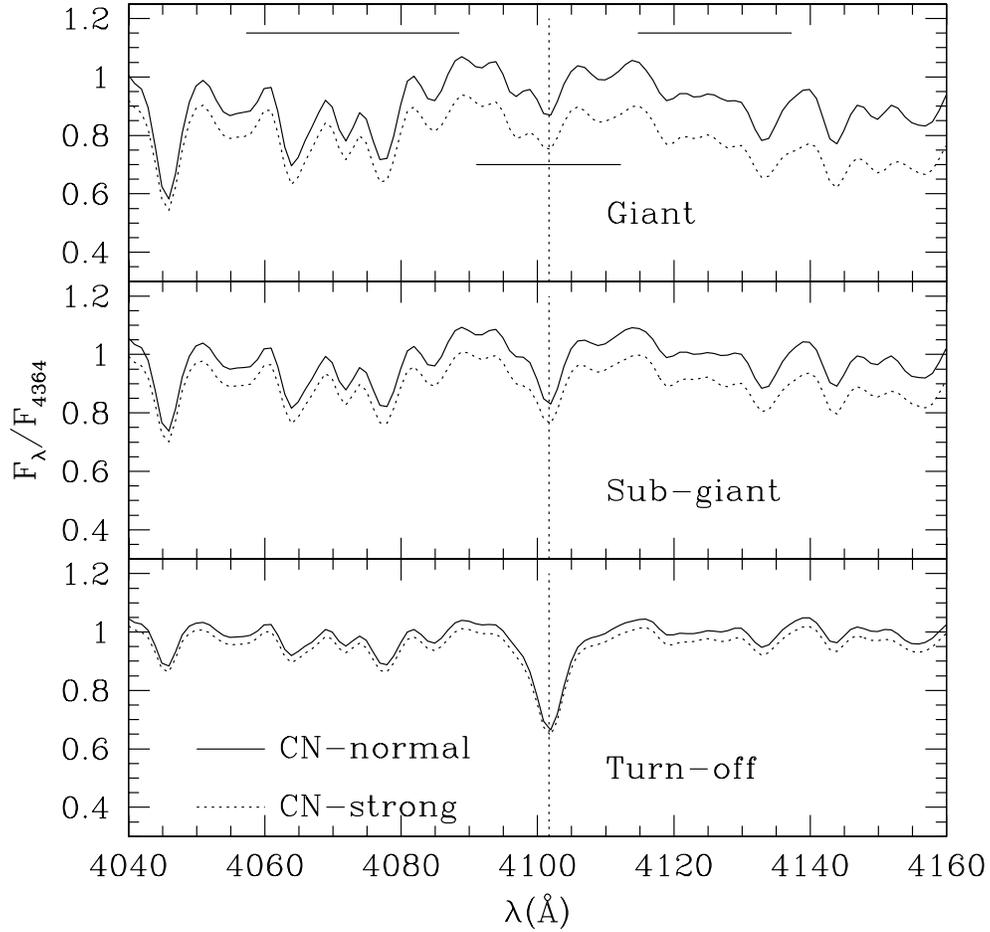}
\caption{Same as Figure \protect\ref{fig13}, in the region of $H\delta$.
The upper panel shows the passbands of the $H\delta_F$ index.
The vertical line indicates the central wavelength of $H\delta$.
CN lines contaminate both the continuum and central bandpasses of
the $H\delta_F$ index, making the final behaviour of the index
very hard to predict. See discussion in the text.
\protect\label{fig14b}
}
\end{figure}

On the basis of the good general agreement between our model predictions
and the observed integrated spectrophotometry and color of the cluster,
we report in Table~\ref{tbl-3} the relative contributions of the various
evolutionary stages to the integrated light of the cluster, at a number
of reference wavelengths. The different evolutionary stages are delimited
in the CMD of Figure \ref{fig4}. We notice that those definitions are
somewhat arbitrary (for instance, in the case of the line separating
turn-off from subgiant stars); therefore the results are subject to
small changes if alternative definitions are adopted. The giants are
divided into 4 sub-classes according to their $T_{eff}$s: G50 ($T_{eff}
\simgreat$ 5000 K), G45 (5000 $\simgreat$ $T_{eff} \simgreat$ 4500 K),
G40 (4500 $\simgreat$ $T_{eff} \simgreat$ 4000 K) and G30 ($T_{eff}
\simless$ 4000 K).

From this exercise, we confirm the well known fact that giants are the
dominant source of light in the optical, being responsible for more than
half the integrated light for $\lambda \simgreat$ 5000 {\AA}. Furthermore,
the red giants in a narrow range of $T_{eff}$ (G45) account for half
of all giant light. In contrast, turn-off stars contribute only about
1/5 of the integrated light at $\sim$4000 {\AA}, and their contribution
drops to half that value around 5300 {\AA}. Thus, when coupled with the
slow evolution of the turn-off at old ages, it is not surprising that
hydrogen lines such as $H\beta$ in the integrated spectra of old stellar
populations change so slowly with age.

Another interesting finding is the very low contribution of blue
stragglers to the integrated light. As the CMD of Figure \ref{fig4}
samples the core of 47 Tuc, the relative number of blue stragglers is
maximized, yet they contribute $\simless$ 1\% of the integrated light
beyond 4000 {\AA}. Such a small contribution was also inferred from
the integrated CaII index of 47 Tuc by Rose (1994). The effect of blue
stragglers on line indices is also negligible, except for $H\delta_F$
and, to a lesser extent, $H\gamma_{\sigma<130}$ and $H\beta$. The
impact of blue stragglers on age determinations from these indices is
discussed in Paper II.

The CMD of Figure \ref{fig4} does not include white dwarfs and
UV-excess stars which were shown to be abundant in the core of 47 Tuc by
Ferraro et al. (2001). In their Figures 1 and 2, however, it can be seen
that those stars, though being almost as numerous as blue stragglers,
are on average 4--5 magnitudes fainter at $m_{F439W}$, so that their
neglect in our computations has no impact on the final result.

The stars just above the subgiant branch in Figure ~\ref{fig4} are due
to crowding and unresolved binaries. They were taken into account in
the calculations by dividing by two their observed magnitudes. Their
contribution to the integrated light ($<$ 0.3 \% at all wavelengths)
and line indices is negligible.

\section{Conclusions}

In this paper, we synthesized the integrated spectrum of 47 Tuc directly
from its color-magnitude diagram. Our models are based on a moderately
high-resolution spectral library and a deep observed CMD of the cluster.

Our CMD-based model spectrum provides a good match to a number of
Balmer line and metal-line indices. Excellent fits are achieved for
$Mg\,b$, $<Fe>$, Ca4227 and $H\beta$. We also obtain satisfactory fits
for Fe4383 and $H\gamma_{\sigma<130}$. The remaining 0.04 {\AA}
mismatch in Fe4383 (4 $\sigma$, see Table 4) is negligible if compared
to the response of the index to {\it [Fe/H]}  ($\sim$ 0.3 {\AA}/dex,
see Vazdekis et al. 2001). Only for $H\delta_F$ do the models fail in
predicting the observed value of the index, but the CN correction for
this index is perhaps the most uncertain.

A number of key factors were responsible for the agreement achieved:

-- We had to correct the model predictions for a limitation of the
spectral library, namely, that it lacks a sufficient number of metal-poor
giants.

-- A small adjustment (by --0.05 dex) was made to the {\it [Fe/H]}
of 47 Tuc. It is not possible to decide if this really reflects
a lower Fe abundance than the Carretta \& Gratton (1997) value, or if
it is due to an error in the zeropoint of the {\it [Fe/H]}-scale of the
library stars. Further evidence for a lower Fe abundance in 47 Tuc, from
a classical detailed abundance analysis comes from Brown \& Wallerstein
(1992), who found {\it [Fe/H]}=--0.8. This issue needs to be addressed
by future high-resolution studies of individual cluster stars. Most
importantly, the --0.05 shift is comfortably within the errorbars
of both our {\it [Fe/H]}-scale and the abundance determinations of
individual stars.

-- The model predictions were corrected for the effect of CN-strong
stars in 47 Tuc. Since our stellar library lacks CN-strong stars, we
resorted to a spectrum synthesis approach. This correction is crucial
to bring the models into agreement with the observations in the case of
Ca4227. The effect on $H\gamma_{\sigma<130}$ is small, and is probably
small as well in the case of Fe4383.  It is also computed to be small
in the case of $H\delta_F$, in disagreement with the observations, but
the spectrum around $H\delta_F$ is heavily blended by CN lines and the
correction is uncertain.

We devoted special attention to fitting Balmer lines because
they are the key indices for spectroscopic age determination. For
$H\gamma_{\sigma<130}$ and $H\beta$, model predictions agree with the
observations within the errorbars. $H\delta_F$, on the other hand, is
overestimated by 0.4 {\AA}. The latter mismatch translates into an age
overestimate, according to the models to be presented in Paper II, of
$\sim$ 4 Gyrs. Residual mismatches in $H\gamma_{\sigma<130}$ and $H\beta$
translate into age underestimates of $\sim$ 1 and 1.5 Gyrs respectively.

The computation of the integrated spectrum from the CMD of the cluster,
being essentially devoid of inputs from theoretical isochrones,
allowed us to infer model-independent estimates of the contribution
of distinct evolutionary stages to the integrated light at a number of
reference wavelengths. Giant branch and horizontal branch stars together
contribute roughly 50\% of the light at $\lambda \sim 4000$ {\AA}.
In the same spectral region, turn-off stars contribute about 20\% of
the integrated light, while subgiants and lower main sequence stars
contribute roughly equal parts to the remaining 30\%. The contribution
due to giants of course increases towards redder wavelengths. In the
continuum of $H\beta$, turn-off stars contribute less than 15\% of the
integrated light. These numbers are only valid for metallicities around
{\it [Fe/H]}=--0.7. For more metal-rich stellar populations, red-giant
light may become less important in the blue. We call the attention of
the reader to yet another caveat, which refers to the definition of the
evolutionary stages themselves, which is somewhat arbitrary, especially
when drawing a line separating turn-off from subgiant stars. Slightly
different definitions would cause small changes in the numbers given
in Table~\ref{tbl-3}.

We derived a new set of stellar parameters for the spectral library
stars, on the basis of a combination of Str\"omgren photometry and
calibrations from the literature for the dwarfs, and a calibration of
spectral features as a function of stellar parameters for the giants.
On the basis of the latter, we derived a new metallicity-dependent
$(B-V)$ vs. $T_{eff}$ transformation for giant stars. We also assessed
systematic errors in our stellar parameters by comparison with
determinations from other sources. Metallicity and $T_{eff}$-scales can
differ among different authors by as much as 0.1 dex and 70 K for 
giants and 0.05 dex and 50 K for dwarfs. A more thorough discussion
of our determinations and the full list of stellar parameters will be
presented elsewhere (Schiavon 2002, in preparation).

The main conclusion of this paper refers to our ability to correctly
reproduce the integrated spectrum and color of a well-known globular
cluster on the basis of the simplest possible assumptions. Our procedure
relies on very few theoretical assumptions and thus allows us to isolate
and study effects due to limitations of the empirical spectral library
and uncertainties in the stellar parameters adopted. We have shown that
such effects are in most cases minor and can be easily corrected. Most
importantly, we conclude that the exceedingly high spectroscopic ages
found for 47 Tuc in previous work ($>$ 20 Gyrs, Gibson et al. 1999)
cannot be attributed either to features of the spectral library or
the calibrations adopted given that the age differences implied by
the residual mismatch in Balmer line EW. The corrections inferred in
this paper are applied in Paper II to our computations of integrated
spectra based on theoretical isochrones.  There, the main goal is to
fit the integrated spectrum of 47 Tuc to determine its absolute age
and metallicity.

\acknowledgments
We would like to thank S. Covino for making available the red integrated
spectrum of 47 Tuc. Raja Guhathakurta is thanked for providing the
HST color-magnitude diagram of 47 Tuc. We thank Kristi Concannon for
making available some of her fitting functions for Balmer line indices
in advance of publication. We would also like to thank A. Vazdekis and
Nicolas Cardiel for helpful discussions. The referee, Brad Gibson,
is thanked for valuable suggestions that substantially improved
this article. R.P.S. thanks the hospitality of the Dept. of Physics
and Astronomy at the University of North Carolina, Chapel Hill, where
part of this work was developed. Likewise, J.A.R. thanks the Astronomy
Department at UC, Santa Cruz for hospitality during a visit in which part
of this work was developed.  This work has made extensive use of the
Simbad database. R.P.S. acknowledges support provided by the National
Science Foundation through grant GF-1002-99 and from the Association
of Universities for Research in Astronomy, Inc., under NSF cooperative
agreement AST 96-13615, and CNPq/Brazil, for financial help in the form of
a travel grant (200510/99-1). This research has also been supported by NSF
grant AST-9900720 to the University of North Carolina, and by NSF grants
AST-9529098 and AST-0071198 to the University of California, Santa Cruz.

\appendix

\section{APPENDIX}

\subsection{Spectrum synthesis from model atmospheres} \label{apsynth}

Synthetic spectra were computed for a set of representative stellar
parameters, typical of giants, subgiants and turn-off stars from 47
Tuc. The stellar parameters adopted were: ($T_{eff}$,$\log g$) = (4750
K,2.50) for a red giant, (6000 K,4.0) for a turn-off star and (5250 K,3.0)
for a subgiant. All spectra were computed for {\it [Fe/H]}=--0.5.

An updated version of the code by Spite (1967), extended to include
molecular lines by Barbuy (1982), where LTE is assumed, was used for
the spectrum synthesis calculations. This code (FSYNTH) now includes
atomic and molecular lines from the UV to near-IR, and the computations
are made faster.

The synthetic spectra were calculated using two different abundance
patterns: one adopting the solar chemical composition (Grevesse \& Sauval,
1998), scaled to the adopted metallicity of --0.5 dex, and the other for
the abundance pattern of CN-strong stars, {\it [C/Fe]}=--0.2 and {\it
[N/Fe]}=+0.8 (Cannon et al. 1998). The ATLAS9 photospheric models, by
Kurucz (1992) were employed. The spectra were calculated in 0.02 {\AA}
steps and later convolved and rebinned to match the resolution and
dispersion of the observed spectra.

The concentration of CN in the stellar photosphere of metal-poor stars
is known to be important in determining the temperature stratification,
by introducing a backwarming effect (see, for instance, Gustafsson
et al. 1975, Drake, Plez \& Smith 1993, Hill et al. 2000). Therefore,
in order to be fully consistent, we would need to use a set of model
photospheres computed with the abundance pattern characteristic
of CN-strong stars. Unfortunately, such model photospheres are not
presently available. However, as shown by Drake, Plez \& Smith (1993),
the temperature change at low optical depths due to varying the CN
concentration by the amounts discussed here is around 75 K for models with
{\it [Fe/H]}=--1.0, being lower for higher metallicity models. Therefore
the computations shown here serve as a first-order approximation to
the problem.

Our list of atomic lines is based on the list of lines identified in
the solar spectrum by Moore et al. (1966) and includes the updated
list and accurate data for Fe I by Nave et al. (1994). Oscillator
strengths for atomic lines were adopted from Nave et al. (1994), Wiese
et al. (1969), Fuhr et al. (1988) and Martin et al. (1988) whenever
available. Otherwise they were obtained by fitting the solar spectrum
(Castilho 1999). The damping constants for the neutral element lines
were calculated using tables of cross sections by Barklem et al. (1998
and references therein), and for the other lines they were determined
by fitting the solar spectrum (Castilho 1999).

Molecular lines from the following electronic systems were taken into
account in the calculations for the $\lambda\lambda$ 4000-4400 {\AA}
interval: MgH (A$^2 \Pi$-X$^2 \Sigma$), C$_2$ (A$^3 \Pi$-X$^3 \Pi$), CN
blue (B$^2 \Sigma$-X$^2 \Sigma$), CH (A$^2 \Delta$-X$^2 \Pi$), CH (B$^2
\Delta$-X$^2 \Pi$), CH (C$^2 \Sigma$-X$^2 \Pi$), OH (A$^2 \Sigma$-X$^2
\Pi$), NH (A$^3 \Pi$-X$^3 \Sigma$) (see Castilho 1999 and Barbuy et al.
2001 for details). Franck-Condon factors with dependence on the rotational
quantum number J as given in Dwivedi et al. (1978) and Bell et al. (1979)
were computed and adopted when possible. For vibrational bands for which
such values were not available, we adopted a constant value kindly made
available to us through computations by P. D. Singh (unpublished).

\begin{figure}[htpb]
\plotone{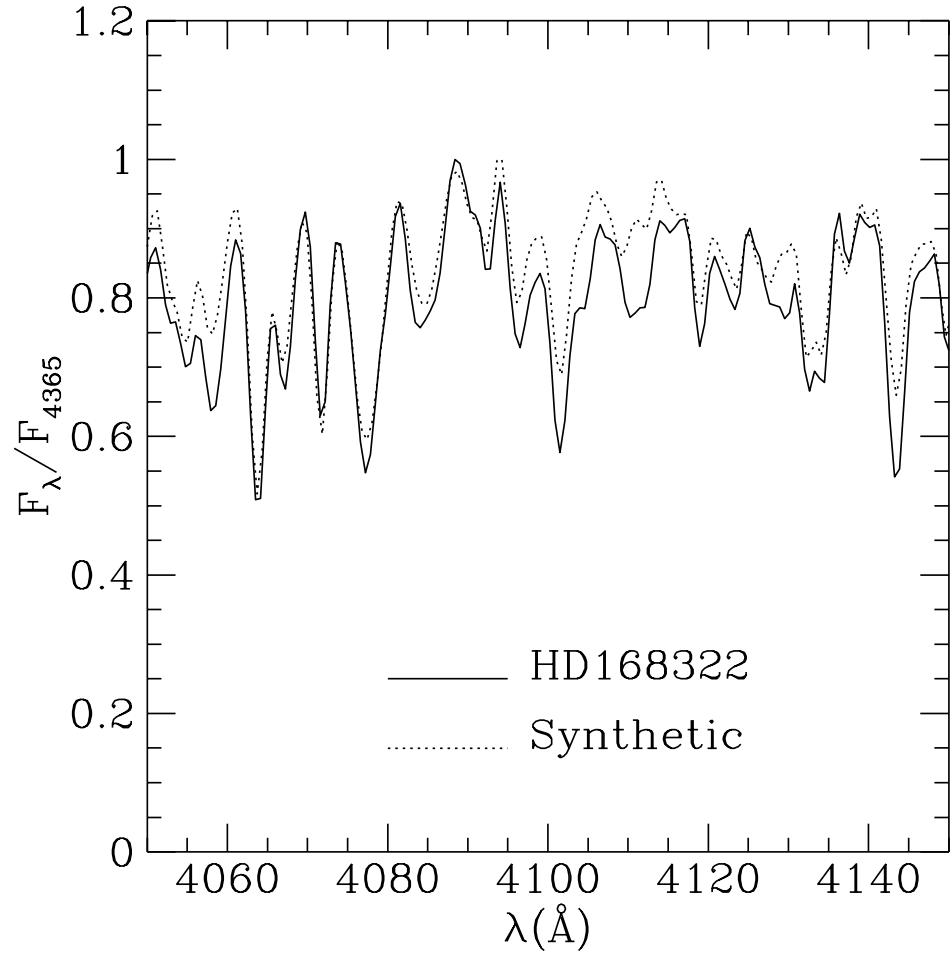}
\caption{Spectrum synthesis in the region of $H\delta$. The star
has the following atmospheric parameters: ($T_{eff}$,$\log
g$,{\it [Fe/H]}) = (4766 K,2.5,-0.5). The synthetic spectrum was computed for
(4750 K,2.5,-0.5).
\protect\label{fig11}
}
\end{figure}

\begin{figure}[htpb]
\plotone{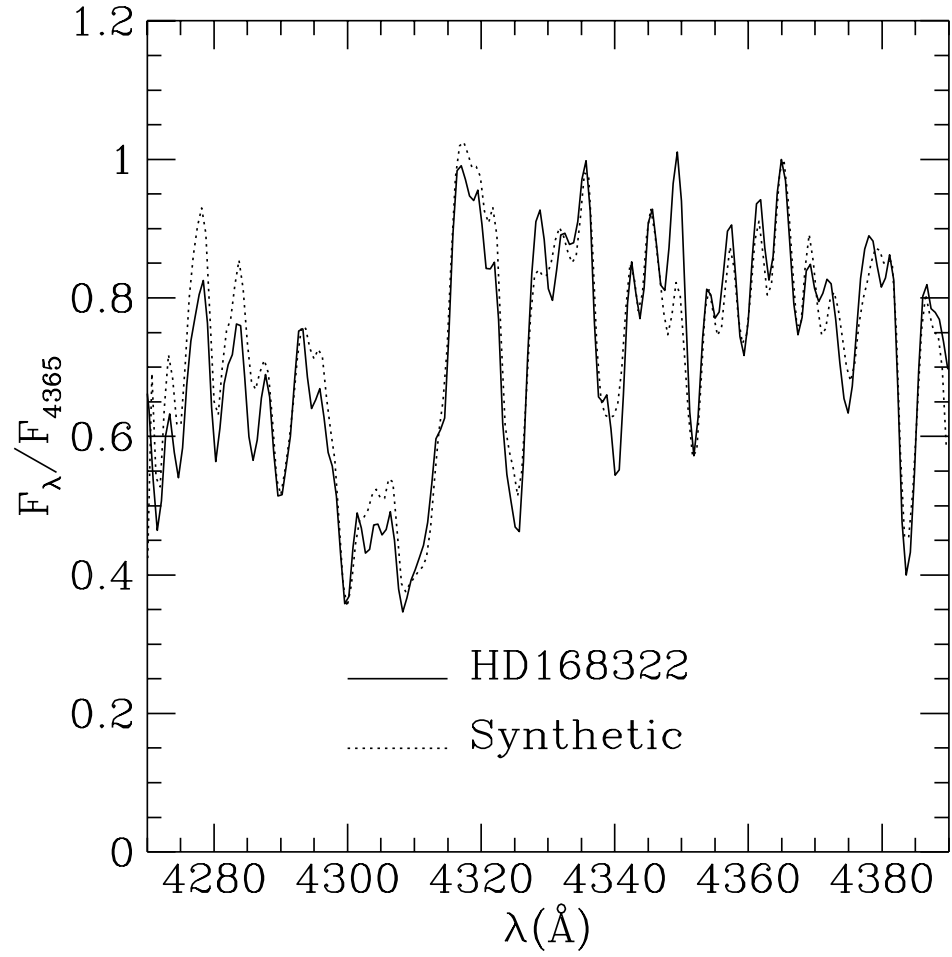}
\caption{Same as Figure \protect\ref{fig11}, in the $H\gamma$ region.
\protect\label{fig12}
}
\end{figure}

Figures \ref{fig11} and \ref{fig12} compare the broadened and
rebinned synthetic spectrum, computed adopting solar abundance ratios,
to the library spectrum of a star with similar stellar parameters in
the neighborhood of both $H\gamma$ and $H\delta$. While most spectral
features are well reproduced by the synthetic spectrum, it is clear
that some lines in the observed spectrum are not well matched by the
models.  The region of biggest mismatch is in fact the line passband of
$H\delta_F$, which might explain our failure to fully match this index
(Section~\ref{lind}).  This is a result of uncertainties in both the
model atmospheres and atomic oscillator strengths and also probable
incompleteness in the atomic line lists.

\clearpage

\clearpage

\clearpage

\clearpage

\clearpage

\clearpage

\clearpage

\clearpage

\clearpage

\clearpage

\clearpage

\clearpage

\clearpage

\clearpage

\clearpage

\clearpage

\clearpage

\clearpage

\clearpage

\begin{deluxetable}{cccccccc}
\tablecaption{Coefficients of the polynomial fits of equations (1) and (2) 
\label{tbl-1}}
\tablewidth{0pt}
\tablehead{
\colhead {} & $a_0\,/\,b_0$ & $a_1\,/\,b_1$ & $a_2\,/\,b_2$ & $a_3\,/\,b_3$ & 
$a_4\,/\,b_4$ & $a_5\,/\,b_5$ & $b_6$ }
\startdata
{\it [Fe/H]} & -2.6895 & 0.47814 & -0.01916 & -0.06726 & 0.75866 & -0.00383 & \\
$T_{eff}$ & 3989.76 & -17.480 & -4.032 & -74.572 & 1072.59 & -118.957 & 
49.325 \\
 \enddata
\end{deluxetable}
\clearpage

\begin{deluxetable}{cccccc}
\tablecaption{Coefficients of the polynomial fit of equation (3) 
\label{tbl-2a}}
\tablewidth{0pt}
\tablehead{
\colhead {} $a_0$ & $a_1$ & $a_2$ & $a_3$ & $a_4$ & $a_5$ }
\startdata
0.70816 & 0.31406 & 0.03296 & 0.09548 & -0.17799 & -0.04965 \\
\enddata
\end{deluxetable}
\clearpage

\begin{deluxetable}{cccc}
\tablecaption{Observed and Predicted $(B-V)_0$ for 47 Tuc \label{tbl-2}}
\tablewidth{0pt}
\tablehead{
\colhead{$(B-V)_0$} & {$(B-V)_{CMD}$} & $(B-V)^{st}_{CMD}$ }
\startdata
0.86$\pm$0.01 & 0.86 & 0.85 \\
\enddata

\tablecomments{$(B-V)_0$ is the dereddened $(B-V)$ from Chun \& Freeman
(1979), within an aperture of 1 arcmin. Superscript `st' denotes the
integrated color of the stars entering the synthesis, while no superscript
indicates the color obtained from direct integration of stars in the
CMD. In both cases, stars below the completeness limit are included
assuming a Salpeter IMF (see text).}

\end{deluxetable}
\clearpage

\begin{deluxetable}{lccccccc}
\tablecaption{Observed and Predicted Line Indices for 47 Tuc \label{tbl-4}}
\rotate
\tablewidth{0pt}
\tablehead{
\colhead {} & $H\delta_F$ & Ca4227 & $H\gamma_{\sigma<130}$ & Fe4383 & $H\beta$ & $\log Mg\,b$ & $<Fe>$   }
\startdata
Observed & 0.777$\pm$0.003 & 0.92$\pm$0.02 & 0.92$\pm$0.04 & 3.16$\pm$0.01 &
1.57$\pm$0.17 & 0.45$\pm$0.03 & 2.29$\pm$0.25  \\
Observed - BS & 0.70 & 0.92 & 0.91 & 3.19 & 1.54 & 0.45 & 2.29 \\
Model    & 0.796 & 1.288 & 0.97 &  3.56 & 1.55 & 0.47 & 2.32 \\
{$[Fe/H]-$}Corr. & 1.06 & 1.26 & 0.93 &  3.20 & 1.48 & 0.46 & 2.14 \\
{CN-}Corr. & 1.17 & 0.80 & 0.89 &  -- & -- & -- & -- \\
\enddata

\tablecomments{The first row lists indices measured in the observed
spectra.  In the second row, the indices are corrected from the
contribution by blue stragglers. The corrections are larger than the
errorbars only for $H\delta_F$ and Fe4383. These are the values to be
compared with predictions from model isochrones in Paper II, while the
CMD-based models listed in rows 3 through 5 should be compared to the
values of the first row. The third row lists the indices measured in
the CMD-based model spectra. The fourth row lists model predictions
corrected for the metallicity effects discussed in Section~\ref{lind}
and fifth row lists the indices corrected for the contamination of Cn
and CH lines, according to the discussion of Section~\ref{synth}.}

\end{deluxetable}
\clearpage

\begin{deluxetable}{ccccccccccc}
\tablecaption{Fractional contribution (\%) to the integrated light from 
stars in different evolutionary
stages, defined in Figure 10 (PMS = G50 + G45 + G40 + G30 +HB). \label{tbl-3}}
\tablewidth{0pt}
\tablehead{
\colhead {$\lambda$} & TO & G50 & G45 & G40 & G30 & HB & PMS & MS & SG & BS }
\startdata
 3912 & 23.5 & 7.6 & 15.0 &  4.3 &  2.4 & 18.6 & 14.4 & 47.9 & 12.5 & 1.4  \\ 
 4088 & 18.8 & 7.9 & 18.0 &  7.0 &  4.4 & 18.7 & 13.4 & 56.0 & 10.5 & 0.8  \\ 
 4317 & 16.9 & 8.1 & 19.1 &  8.6 &  5.7 & 19.0 & 12.4 & 60.5 & 9.3 &  0.7  \\ 
 4475 & 15.5 & 8.1 & 19.7 &  9.7 &  6.5 & 18.8 & 12.1 & 62.8 & 8.7 &  0.6  \\ 
 4896 & 13.6 & 7.6 & 20.4 & 11.8 &  9.7 & 17.2 & 11.3 & 66.7 & 7.7 &  0.5  \\ 
 5358 & 12.2 & 7.2 & 20.5 & 13.4 & 11.6 & 17.3 & 10.3 & 70.0 & 6.8 &  0.4  \\ 

\enddata
\end{deluxetable}

\end{document}